\documentclass[prd,reprint,groupedaddress,nofootinbib]{revtex4-2}

\usepackage{graphicx}
\usepackage{dcolumn}
\usepackage{bm}
\usepackage{mathtools}
\usepackage{amssymb,amsmath}
\usepackage[protrusion=true,expansion=true]{microtype}
\usepackage[dvipsnames]{xcolor}
\usepackage{microtype}
\usepackage{siunitx}
\usepackage{multirow}
\usepackage{dsfont}
\usepackage{slashed}
\usepackage[colorlinks=true,linkcolor=blue,citecolor=blue,urlcolor=blue]{hyperref}

\usepackage{ragged2e}
\newcolumntype{L}[1]{>{\raggedright\arraybackslash}p{#1}}
\newcolumntype{C}[1]{>{\centering\arraybackslash}p{#1}}
\newcolumntype{R}[1]{>{\raggedleft\arraybackslash}p{#1}}
\newcolumntype{J}[1]{>{\justifying\arraybackslash}p{#1}}

\newcommand{\vect}[1]{{\mbox{\boldmath $#1$}}}

\definecolor{webred}{rgb}{0.75,0,0}

\def\mA{\ensuremath{\mathcal{A}}}
\def\mC{\ensuremath{\mathcal{C}}}

\def\mM{\ensuremath{\mathcal{M}}}
\def\mN{\ensuremath{\mathcal{N}}}

\def\mQ{\ensuremath{\mathcal{Q}}}

\def\mS{\ensuremath{\mathcal{S}}}

\def\mV{\ensuremath{\mathcal{V}}}
\def\mW{\ensuremath{\mathcal{W}}}

\def\pT{\ensuremath{\mathsf{T}}}
\def\pC{\ensuremath{\mathsf{C}}}
\def\pP{\ensuremath{\mathsf{P}}}
        
\def\sS{\ensuremath{\mathbf{1}}}        
\def\aA{\ensuremath{\mathbf{\overline{1}}}} 
\def\sQ{\ensuremath{\mathbf{4}}}        
\def\aQ{\ensuremath{\mathbf{\overline{4}}}} 
\def\sV{\ensuremath{\mathbf{5}}}        
\def\aV{\ensuremath{\mathbf{\overline{5}}}} 
\def\sW{\ensuremath{\mathbf{6}}}

\DeclareMathOperator{\Tr}{Tr}

\usepackage[vcentermath]{youngtab}

\begin{document}

\title{Five-point functions and the permutation group $S_5$}
 
\author{Gernot Eichmann$^1$}
\email[]{gernot.eichmann@uni-graz.at}
\author{Raul D. Torres$^{2,3}$}
\email[]{raul.torres@tecnico.ulisboa.pt}
\affiliation{$^1$Institute of Physics, University of Graz, NAWI Graz, Universitätsplatz 5, 8010 Graz, Austria}
\affiliation{$^2$Departamento de Física, Instituto Superior Técnico, Universidade de Lisboa, Av. Rovisco Pais 1, 1049-001 Lisboa, Portugal}
\affiliation{$^3$Laboratório de Instrumentação e Física Experimental de Partículas, Av. Prof. Gama Pinto 2, 1649-003 Lisboa, Portugal}

\date{\today}

\begin{abstract}
Five-point functions and five-body wave functions play an important role in many areas of nuclear and particle physics, 
e.g., in $2\to 3$ scattering processes, in the five-gluon vertex, or in the study of pentaquarks.
In this work we consider the permutation group $S_5$ to facilitate the description of such objects.
We work out the multiplets transforming under irreducible representations of $S_5$ 
and  provide compact formulas allowing one to cast the permutations of an object $f_{12345}$
into combinations with definite permutation symmetry. 
We also give the explicit expressions for the irreducible multiplet products.
We consider several practical applications as examples: 
We arrange the four-momenta and  Lorentz invariants of a five-point function into the multiplet structure,
we work out the color tensors of the five-gluon vertex in the multiplet notation,
and we discuss  applications for five-body wave functions like those of pentaquarks. 
\end{abstract}

\maketitle

\section{Introduction}\label{intro}

Permutation-group symmetries are a powerful tool in physics.
If a system is invariant under permutations, then it is a singlet under the respective permutation group $S_n$.
The paradigmatic example in hadron physics is the wave function of a baryon in the quark model:
For three identical quarks, it must be totally antisymmetric under the exchange of any two quarks,
so it is an antisinglet under $S_3$. Its individual parts -- spatial, spin, flavor and color --
can then be grouped into multiplets transforming under irreducible representations of $S_3$,
such that their product yields an antisinglet.

The primary quantities of interest in quantum field theory are its $n$-point correlation functions.
Their generic form is 
 \begin{equation}\label{npt-fct-intro}
   \begin{split}
      \Gamma^{\mu\nu\cdots}_{\alpha\beta\cdots}(p_1\cdots p_n)= 
      \sum_{i} f_i(\dots)(\tau_i)^{\mu\nu\cdots}_{\alpha\beta\cdots}(p_1 \cdots p_n)\,, 
   \end{split}
\end{equation}
where the $p_1 \cdots p_n$ are the four-momenta in the system. 
The $\tau_i$ are the Lorentz-covariant tensor basis elements, with Lorentz (Dirac) indices as superscripts (subscripts), 
and possibly further parts representing flavor, color, etc.
The coefficients $f_i$ are Lorentz-invariant dressing functions, which depend
on the invariant momentum variables and encode the dynamics of the process. 
The structure of $n$-body wave function is analogous to Eq.~\eqref{npt-fct-intro}
except that the sum $p_1 + \dots + p_n$ is not zero but equal to the momentum $P$ of the bound state or resonance.

Clearly, for a growing number of external legs the increasingly complex structure in Eq.~\eqref{npt-fct-intro}
becomes difficult to manage. 
Here, permutation-group symmetries turn out to be extremely useful.
If one casts the $\tau_i$ into $S_n$ multiplets  
such that they share the permutation symmetry of the $n$-point function,
the resulting $f_i$ are singlets. As such, they often show a `planar degeneracy', i.e., they mainly depend
on the symmetric variable $p_1^2 + \dots + p_n^2$  while the dependence on the remaining variables
is small or even negligible. This behavior is found in
the three-gluon vertex~\cite{Eichmann:2014xya,Pinto-Gomez:2022brg,Ferreira:2023fva,Aguilar:2023qqd}, 
the four-gluon vertex~\cite{Aguilar:2024fen}, 
the hadronic light-by-light scattering amplitude~\cite{Eichmann:2015nra,Eichmann:2014ooa},
in two-photon transition form factors~\cite{Eichmann:2017wil,Eichmann:2024glq}, nucleon Compton scattering~\cite{Eichmann:2018ytt} and other examples.
Similarly, in the calculation of three- and four-body hadronic wave functions, one may switch off entire multiplets
of variables  without significant loss of information~\cite{Eichmann:2015cra}.
Even if a system is not symmetric, the permutation-group methods can still be applied and are a useful starting point for quantifying deviations.

In the present work we extend a similar analysis of the permutation group $S_4$~\cite{Eichmann:2015nra} to five-body systems.
Five-body amplitudes appear in many different contexts, e.g., 
when using functional methods to calculate the $n$-point functions of Quantum Chromodynamics (QCD)~\cite{Eichmann:2016yit,Huber:2018ned,Huber:2020keu,Dupuis:2020fhh,Cyrol:2017ewj} 
or quantum gravity~\cite{Meibohm:2015twa,Bonanno:2021squ};
in the description of scattering amplitudes at high energies~\cite{DelDuca:2009ae,Badger:2017jhb,Buccioni:2024gzo}
and their general properties using various different techniques~\cite{Sjodahl:2008fz,Ahmadiniaz:2016qwn,Chiodaroli:2017ngp,Chu:2023kpe,Ochirov:2016ewn,Keppeler:2023msu,Miesch:2024vjk};
in studies of exotic pentaquark states observed at LHCb \cite{LHCb:2015yax,LHCb:2019kea,LHCb:2022ogu};
for amplitude analyses of  $2\to 3$ reactions such as $\pi N\to \pi\pi N$, $\gamma N\to \pi\pi N$ 
or $\pi N \to \eta^{(')} \pi N$~\cite{Byckling-Kajantie,Mokeev:2015lda,JPAC:2018zyd}
and their treatment using lattice QCD and effective field theories \cite{Aaron:1966zz,Aaron:1968aoz,Briceno:2017tce,Mai:2017vot,Jackura:2020bsk,Mai:2022eur,Severt:2022jtg};
or in the form of the $3\to 2$ annihilation process in  strongly interacting dark matter scenarios \cite{Hochberg:2014dra,Hochberg:2014kqa},
which can be effectively described by the Wess-Zumino-Witten term~\cite{Wess:1971yu,Witten:1983tw}.
 
The paper is organized as follows. In Sec.~\ref{s5} we derive the $S_5$ multiplets and multiplet products. 
In Sec.~\ref{5pa} we study as examples the momentum dependence of five-point functions, the color structure of the five-gluon vertex,
and the kinematic dependencies in  pentaquark Bethe-Salpeter amplitudes. We summarize in Sec.~\ref{summary}, and
the appendices collect technical details.

\begin{figure*}[t]
   \begin{center}
   \includegraphics[width=1\textwidth]{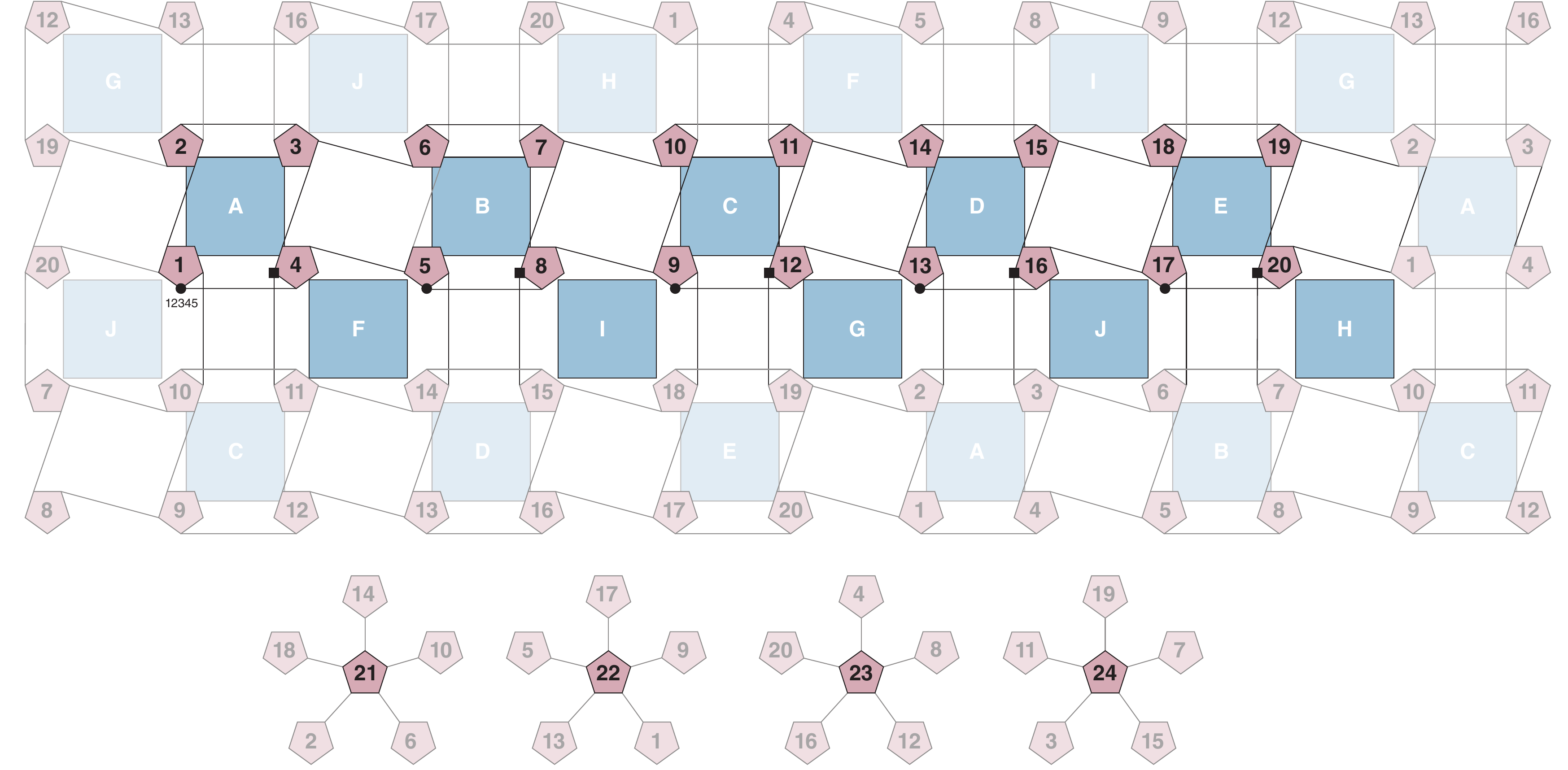}
   \caption{A possible representation of the Cayley graph for the permutation group $S_5$.}
            \label{fig:cayley}
   \end{center}
   \vspace{-3mm}
\end{figure*}

\section{Permutation group $S_5$}\label{s5}

\subsection{Cayley graph}

The permutation group $S_5$ consists of $5!=120$ elements. Each permutation of an object $f_{12345}$ can
be reconstructed from two group elements, a transposition $\pT = P_{12}$ and a 5-cycle $\pC = P_{12345} = P_{12}\,P_{23}\,P_{34}\,P_{45}$. The former
interchanges the indices $1\leftrightarrow 2$ and the latter
is a cyclic permutation $1\rightarrow 2$, $2\rightarrow 3$, $3\rightarrow 4$, $4\rightarrow 5$, $5\rightarrow 1$. For example, one has
\begin{equation}     
   \begin{split}
      \pC\,f_{35124} &= P_{12}\,P_{23}\,P_{34}\,P_{45}\,f_{35124}  \\
                     &= P_{12}\,P_{23}\,P_{34}\,f_{34125}  \\
                     &= P_{12}\,P_{23}\,f_{43125} \\
                     &= P_{12}\,f_{42135} = f_{41235}\,.
   \end{split}
\end{equation}
        
The group manifold of $S_5$ is represented by its Cayley graph.
Because every permutation can be reconstructed from transpositions and cyclic permutations,
the Cayley graph can be visualized by 120/5 = 24 pentagons, which are connected to each other by transpositions. 
The corners of each pentagon 
correspond to the five possible cyclic permutations, which we choose to run in clockwise direction. 
Each corner is linked to another 
pentagon by a transposition $\pT$. 
        
A possible representation of the Cayley graph is shown in Fig.~\ref{fig:cayley}: 
The  figure on the top contains 20 pentagons \mbox{(\#1 \dots \#20)}, whose four corners are connected 
to other pentagons by transpositions and thereby form a plane.
The remaining corner element in each pentagon is connected to the pentagons \#21 \dots \#24 at the bottom of the figure by another transposition.
In this way, the 120 permutations are divided into ten subsets, whose members are obtained by two successive chains of 
the form $( \pC^2, \pT, \pC^2, \pT, \pC \pT, \pT)$, e.g.:
\begin{equation}\label{cayley-path}
   \begin{split}
      f_{12345} \, & \stackrel{\pC^2}{\longrightarrow} \, f_{34512} \, \stackrel{\pT}{\longrightarrow} \, f_{34521}  
                     \stackrel{\pC^2}{\longrightarrow} \, f_{51243} \, \stackrel{\pT}{\longrightarrow} \, f_{52143} \\
                   & \stackrel{\pC \pT}{\longrightarrow} \, f_{12354} \, \stackrel{\pT}{\longrightarrow} \, f_{21354} \\
                   & \stackrel{\pC^2}{\longrightarrow} \, f_{43521} \, \stackrel{\pT}{\longrightarrow} \, f_{43512}  
                     \stackrel{\pC^2}{\longrightarrow} \, f_{15234} \, \stackrel{\pT}{\longrightarrow} \, f_{25134} \\
                   & \stackrel{\pC \pT}{\longrightarrow} \, f_{21345} \, \stackrel{\pT}{\longrightarrow} \, f_{12345}\,.\\
   \end{split}
\end{equation}
Note that $\pC \pT$ implies performing a transposition first and a cyclic permutation afterwards.
The extra $\pT \to \pC \pT$ step ensures that also all elements in the 
remaining four pentagons \#21 \dots \#24 are included.
The resulting path is the square  `A' in Fig.~\ref{fig:cayley}. 

In total, this leads to ten  squares, where each square defines a closed path with twelve elements 
like in Eq.~\eqref{cayley-path}.
For later use, we collect the elements in the first five squares (A, B, C, D, E) into five vectors $f^{(n)}$ with $n=1 \dots 5$:
\begin{equation*}
		\left[
			\begin{array}{c}
				f_{12345} \\ f_{34512} \\ f_{52143} \\ f_{21354} \\ f_{43521} \\ f_{25134} 				
			\end{array}
		\right]\!, 
		\left[
			\begin{array}{c}
				f_{43152} \\  f_{15324} \\  f_{41532} \\  f_{53241} \\  f_{25413} \\  f_{31245}  				
			\end{array}
		\right]\!, 
        \left[
			\begin{array}{c}
				f_{51423} \\ f_{23145} \\ f_{35421} \\ f_{42513} \\ f_{14235} \\ f_{42351} 				
			\end{array}
		\right]\!, 
		\left[ 
			\begin{array}{c}
				f_{24531} \\ f_{41253} \\ f_{24315} \\ f_{15432} \\ f_{32154} \\ f_{53412} 				
			\end{array}
		\right]\!, 
		\left[
			\begin{array}{c}
				f_{35214} \\ f_{52431} \\ f_{13254} \\ f_{34125} \\ f_{51342} \\ f_{14523} 				
			\end{array}
		\right]\!.
  \label{eq:s5 elements}
\end{equation*}   
Together with the transpositions $\pT f^{(n)}$, this yields twelve elements for each vector.
The first entry in each vector is indicated by a black circle in Fig. \ref{fig:cayley}, e.g.,
$f^{(1)}_1 = f_{12345}$ in pentagon \#1, $f^{(2)}_1 = f_{43152}$ in pentagon \#5, and so on.

The remaining five squares (F, G, H, I, J) define the vectors $\tilde f^{(n)}$ with $n=1 \dots 5$:
\begin{equation*}
		\left[
			\begin{array}{c}
				f_{32451} \\ f_{54123} \\ f_{12435} \\ f_{31542} \\ f_{53214} \\ f_{15342}				
			\end{array}
		\right]\!, 
		\left[
			\begin{array}{c}
				f_{13524} \\ f_{35241} \\ f_{51324} \\ f_{23415} \\ f_{45132} \\ f_{21453} 				
			\end{array}
		\right]\!, 
		\left[
			\begin{array}{c}
        f_{41235} \\ f_{13452} \\ f_{45213} \\ f_{52134} \\ f_{24351} \\ f_{32514} 				
			\end{array}
		\right]\!, 
		\left[
			\begin{array}{c}
        f_{54312} \\ f_{21534} \\ f_{34152} \\ f_{45321} \\ f_{12543} \\ f_{43125} 				
			\end{array}
		\right]\!, 
		\left[
			\begin{array}{c}
				f_{25143} \\ f_{42315} \\ f_{23541} \\ f_{14253} \\ f_{31425} \\ f_{54231} 				
			\end{array}
		\right]\!.  
  \label{eq:s5 elements-2}
\end{equation*}    
Again, combined with the transpositions $\pT \tilde f^{(n)}$,
each vector $\tilde f^{(n)}$ defines a closed path with twelve elements.
Here the first entry in each vector is shown by a black square in Fig.~\ref{fig:cayley}, e.g.,
$\tilde f^{(1)}_1 = f_{32451}$ in pentagon \#4 (square F), $\tilde f^{(2)}_1 = f_{13524}$ in pentagon \#12 (square G), and so on.
Denoting the column index by $m=1 \dots 6$, the 120 elements are then defined by 
\begin{equation}
   f_m^{(n)}\,, \quad \pT f^{(n)}_m\,, \quad \tilde f^{(n)}_m\,, \quad \pT \tilde f^{(n)}_m\,.
\end{equation}

\subsection{$S_5$ multiplets}
        
Next, we want to rearrange the 120 permutations into multiplets
that transform under the irreducible representations of $S_5$. 
These are given by the Young diagrams of $S_5$, which we label by bold fonts for brevity: 
\begin{equation*}
   \begin{array}{c @{\quad} c @{\quad} c @{\quad\;\;} c @{\quad\;\;} c @{\quad\;\;} c @{\quad\;\;} c}
      \sS & \sQ & \sV & \sW &  \aV & \aQ & \aA \\[1mm]
      \mS & \mQ^+_i & \mV^+_j & \mW_k &  \mV^-_j & \mQ^-_i & \mA \\[2mm]
      \scriptsize\yng(5) &
      \scriptsize\yng(4,1) &
      \scriptsize\yng(3,2) &
      \scriptsize\yng(3,1,1) &
      \scriptsize\yng(2,2,1) &
      \scriptsize\yng(2,1,1,1) &
      \scriptsize\yng(1,1,1,1,1)
   \end{array}
\end{equation*}
We will denote singlets by $\mS$ and antisinglets by $\mA$.
There are four quartets $\mQ^+_i$ and four antiquartets $\mQ^-_i$ ($i=1 \dots 4$), which transform under inequivalent irreducible representations
and thereby form two different four-dimensional subspaces.
The five quintets and five antiquintets $\mV^\pm_j$  ($j=1 \dots 5$) form two five-dimensional subspaces,
and the six sextets $\mW_k$ ($k=1 \dots 6$) form a six-dimensional subspace.
In total, this gives  $1 + 4^2 + 5^2 + 6^2 +5^2 + 4^2 + 1= 120$ elements.
These are linear combinations of the object $f_{12345}$ and its permutations,
which we want to work out in the following.
        
\renewcommand{\arraystretch}{1.4}
Written as \mbox{four-}, five- and six-dimensional column vectors,
the multiplets transform under the matrix representations of $S_5$.
The transformation laws for any permutation can be reconstructed from a transposition $\pT$ and a cyclic permutation $\pC$. They are given by
\begin{equation}\label{permutation-tf-s4} \renewcommand{\arraystretch}{1.2}
   \begin{split}
      \pT\left[ \begin{array}{c} \mS \\ \mA \\ \mQ_i^\lambda \\[1mm] \mV_j^\lambda \\[1mm] \mW_k \end{array}\right] =
         \left[ \begin{array}{c} \mS \\ -\mA \\ \lambda\,\mathbf{T}_{4}^T\,\mQ_i^\lambda \\[1mm] 
            \lambda\,\mathbf{T}_{5}^T\,\mV_j^\lambda \\[1mm]  \mathbf{T}_{6}^T\,\mW_k \end{array}\right] ,\; 
      \pC\left[ \begin{array}{c} \mS \\ \mA \\ \mQ_i^\lambda \\[1mm] \mV_j^\lambda \\[1mm] \mW_k \end{array}\right] =
         \left[ \begin{array}{c} \mS \\ \mA \\ \mathbf{C}_{4}^T\,\mQ_i^\lambda \\[1mm] 
            \mathbf{C}_{5}^T\,\mV_j^\lambda \\[1mm]  \mathbf{C}_{6}^T\,\mW_k \end{array}\right]. 
        \renewcommand{\arraystretch}{1.0}
   \end{split}
\end{equation}
For example, a singlet is invariant under any permutation ($\pT \,\mS = \mS$, $\pC\,\mS = \mS$).
An antisinglet is invariant under cyclic permutations ($\pC\,\mA = \mA$) but picks up a minus sign under transpositions ($\pT \,\mA = -\mA$).
The remaining multiplets transform under four-, five- or six-dimensional representation matrices, where the superscript $T$ denotes a matrix transpose.
These can be worked out using standard methods~\cite{van1998some}. 
It is always possible to choose the basis vectors in the derivation such that the matrices for the quartets and antiquartets,
as well as those for the quintets and antiquintets, differ only by minus signs, hence the common transformation law in Eq.~\eqref{permutation-tf-s4} with $\lambda=\pm$.
The resulting transposition matrices read
\begin{equation}
   \begin{split}
      \mathbf{T}_4 &= \text{diag}(-1,1,1,1)\,, \\
      \mathbf{T}_5 &= \text{diag}(1,1,-1,-1,1)\,, \\
      \mathbf{T}_6 &= \text{diag}(-1,-1,-1,1,1,1)  
   \end{split}
\end{equation}
and those for the cyclic permutations are given by
\renewcommand{\arraystretch}{1.2}
        \begin{align}\label{triplet-rep-matrices}
   \mathbf{C}_4 &= -\frac{1}{12}\left(\begin{array}{cccc} 6 &  2\sqrt{3}  &   \sqrt{6} & 3\sqrt{10} \\ 
                   -6\sqrt{3} & 2 & \sqrt{2} & \sqrt{30} \\ 
                          0 & -8\sqrt{2} & 1 & \sqrt{15} \\
                         0 & 0 & -3\sqrt{15} & 3 
                                      \end{array}\right),  \nonumber \\
   \mathbf{C}_5  &= -\frac{1}{12}\left(\begin{array}{ccccc} 4 &  -4\sqrt{6}  &  0 & 0 & 4\sqrt{2} \\ 
                       0 & 3 & 3\sqrt{3} & 9 & 3\sqrt{3} \\ 
             0 & 3\sqrt{3} & -3 & -3\sqrt{3} & 9 \\
             4\sqrt{6} & 3 &  3\sqrt{3} & -3 & -\sqrt{3} \\
       4\sqrt{2} & \sqrt{3} & -9 & 3\sqrt{3} & -1 \end{array}\right), \\
   \mathbf{C}_6  &= \frac{1}{24}\left(\begin{array}{cccccc} 8 &  4\sqrt{2}  &  4\sqrt{30} & 0 & 0 & 0 \\ 
            -8\sqrt{2} & 1 & \sqrt{15} & 3\sqrt{3} & 9\sqrt{5} & 0\\ 
             0 & -3\sqrt{15} & 3 & -3\sqrt{5} & \sqrt{3} & 8\sqrt{6} \\
             8\sqrt{6} & -\sqrt{3} &  -3\sqrt{5} & 3 & 3\sqrt{15} & 0 \\
             0 & 9\sqrt{5} & -3\sqrt{3} & -\sqrt{15} & 1 & 8\sqrt{2} \\
             0 & 0 & 0 & 4\sqrt{30} & -4\sqrt{2} & 8 \end{array}\right)\!. \nonumber
\end{align}

Once the representation matrices are known, it is in principle straightforward to work out the explicit forms of the multiplets.
In practice, the task is somewhat tedious since Eq.~\eqref{permutation-tf-s4} amounts to solving large systems of algebraic equations 
(e.g., $6 \times 2 \times 120 = 1440$ equations to obtain the sextets)
and the resulting expressions are rather lengthy.
Here our definitions of the $f^{(n)}_m$ and $\tilde f^{(n)}_m$  turn out to be  useful.
Denoting the (anti-)symmetrizer by $\pP_\pm = 1 \pm \pT = 1 \pm \pP_{12}$,
We abbreviate
\begin{equation}
\begin{split}
   \psi_{1,nm}^{\lambda,\pm} &= \pP_\lambda \left( f^{[n]}_m \pm \tilde f^{[n]}_m \right), \\
   \psi_{2,nm}^{\lambda,\pm} &= \pP_\lambda \left( f^{[n-1]}_m \pm \tilde f^{[n-3]}_m \right), \\
   \psi_{3,nm}^{\lambda,\pm} &= \pP_\lambda \left( f^{[n-3]}_m \pm \tilde f^{[n-4]}_m \right), \\
   \psi_{4,nm}^{\lambda,\pm} &= \pP_\lambda \left( f^{[n-4]}_m \pm \tilde f^{[n-2]}_m \right), \\
   \psi_{5,nm}^{\lambda,\pm} &= \pP_\lambda \left( f^{[n-2]}_m \pm \tilde f^{[n-1]}_m \right), 
\end{split}
\end{equation}
with $\lambda = \pm$ and $[n]=\text{mod}(n,5)$, such that
\begin{equation}
   [n] = \left\{ \begin{array}{ll} n & \quad n=1 \dots 5\,, \\ n+5   & \quad n=-4 \dots 0\,. \end{array}\right.
\end{equation}
The multiplets can then be written in the compact form 
\begin{equation}\label{multiplets-final}
\begin{split}
   \mathcal{S}&=\sum^{5}_{n=1}\sum^{6}_{m=1} \psi_{1,nm}^{++}\,, \\
   \mathcal{A}&=\sum^{5}_{n=1}\sum^{6}_{m=1} \psi_{1,nm}^{--}\,, \\
   \mQ^\pm_i &= \sum^{5}_{n=1}\sum^{6}_{m=1} \left[ \begin{array}{l} 
    a_{nm}^{(1)}\,\psi_{i,nm}^{\mp\pm} \\ 
    a_{nm}^{(2)}\,\psi_{i,nm}^{\pm\pm} \\ 
    a_{nm}^{(3)}\,\psi_{i,nm}^{\pm\pm} \\ 
    a_{nm}^{(4)}\,\psi_{i,nm}^{\pm\pm} 
    \end{array}\right], \quad i= 1\dots 4,\\
   \mV^\pm_j &= \sum^{5}_{n=1}\sum^{6}_{m=1} \left[ \begin{array}{l} 
    b_{nm}^{(1)}\,\psi_{j,nm}^{\pm\pm} \\ 
    b_{nm}^{(2)}\,\psi_{j,nm}^{\pm\pm} \\ 
    b_{nm}^{(3)}\,\psi_{j,nm}^{\mp\pm} \\ 
    b_{nm}^{(4)}\,\psi_{j,nm}^{\mp\pm} \\ 
    b_{nm}^{(5)}\,\psi_{j,nm}^{\pm\pm} 
    \end{array}\right],   \quad j=1 \dots 5,   \\ 
   \mW_k &= \sum^{5}_{n=1}\sum^{6}_{m=1} \left[ \begin{array}{l} 
    c_{nm}^{(1)}\,\psi_{k,nm}^{-+} \\ 
    c_{nm}^{(2)}\,\psi_{k,nm}^{-+} \\ 
    c_{nm}^{(3)}\,\psi_{k,nm}^{-+} \\ 
    c_{nm}^{(4)}\,\psi_{k,nm}^{++} \\ 
    c_{nm}^{(5)}\,\psi_{k,nm}^{++} \\ 
    c_{nm}^{(6)}\,\psi_{k,nm}^{++} 
    \end{array}\right],   \quad k=1 \dots 5,   \\ 
   \mW_6 &= \sum^{5}_{n=1}\sum^{6}_{m=1} \left[ \begin{array}{l} 
    d_{m}^{(1)}\,\psi_{1,nm}^{-+} + e_{m}^{(1)}\,\psi_{1,nm}^{--} \\ 
    d_{m}^{(2)}\,\psi_{1,nm}^{-+} + e_{m}^{(2)}\,\psi_{1,nm}^{--}\\ 
    d_{m}^{(3)}\,\psi_{1,nm}^{-+} + e_{m}^{(3)}\,\psi_{1,nm}^{--}\\ 
    d_{m}^{(4)}\,\psi_{1,nm}^{++} + e_{m}^{(4)}\,\psi_{1,nm}^{+-}\\ 
    d_{m}^{(5)}\,\psi_{1,nm}^{++} + e_{m}^{(5)}\,\psi_{1,nm}^{+-}\\ 
    d_{m}^{(6)}\,\psi_{1,nm}^{++} + e_{m}^{(6)}\,\psi_{1,nm}^{+-}
    \end{array}\right],   \\ 
\end{split}
\end{equation}
where the numbers $a^{(1 \dots 4)}_{nm}$, $b^{(1 \dots 5)}_{nm}$, $c^{(1 \dots 6)}_{nm}$, $d^{(1 \dots 6)}_{m}$ and $e^{(1 \dots 6)}_{m}$ 
are collected in Appendix~\ref{App:multiplets}. 
These expressions satisfy the transformation laws in Eq.~\eqref{permutation-tf-s4},
which can be easily verified, e.g., using \texttt{Mathematica}.
Of course, any linear combination of the four quartets $\mQ^+_i$ is again a quartet, and likewise for the antiquartets, quintets, antiquintets and sextets.

With Eq.~\eqref{multiplets-final} it is straightforward to work out the $S_5$ multiplets from a given  `seed element' $f_{12345}$.
As an example, consider $f_{12345} = f_5$. Its possible permutations appearing in the vectors $f^{(n)}_m$ and $\tilde f^{(n)}_m$ 
are $f_1$, $f_2$, $f_3$ and $f_4$, so altogether there are five independent elements.
These can be arranged in $S_5$ multiplets using the procedure above. One obtains  a singlet and one linearly independent quartet,
\begin{align} 
  \mS &= f_1+f_2+f_3+f_4+f_5\,,  \label{example-1} \\
   \mQ^+ &= \left[ \begin{array}{c} 
                                            f_1 - f_2 \\ 
                                            \frac{1}{\sqrt{3}} \,( f_1 + f_2 -2f_3) \\
                                            \frac{1}{\sqrt{6}} \,( f_1 + f_2 + f_3 - 3f_4) \\
                                            \frac{1}{\sqrt{10}}\,( f_1 + f_2 + f_3 + f_4 - 4f_5 ) \end{array}
                                            \right], \label{example-2}
\end{align}
whereas all other multiplets vanish. Note that the choice of $f_5$ is not relevant, any other seed ($f_1$, $f_2$, $f_3$ or $f_4$)
would have given the same result.

      \begin{table}
    \centering
    \begin{tabular}{l @{\quad\;} c @{\quad} r @{\quad} c @{\quad} r @{\quad} r @{\quad} c } \hline\noalign{\smallskip}
                 & $\sQ \otimes \sQ$   & $\sV \otimes \sV$      & $\sW \otimes \sW$  & $\sQ \otimes \sV$   & $\sQ \otimes \sW$  & $\sV \otimes \sW$         \\ \noalign{\smallskip}\hline\noalign{\smallskip}
    $\sS$        & $\mQ\cdot\mQ'$      & $\mV\cdot\mV'$         & $\mW\cdot\mW'$     &                     &                    &                     \\[0.5mm]
    $\aA$        &                     &                        & $\mW\times\mW'$    &                     &                    &                         \\[0.5mm]
    $\sQ$        & $\mQ\vee\mQ'$       & $\mV\vee\mV'$          & $\mW\vee\mW'$      & $\mQ\vee\mV$        & $\mQ\vee\mW$       & $\mV\vee\mW$             \\[0.5mm]
    $\aQ$        &                     & $\mV\wedge\mV'$        & $\mW\wedge\mW'$    &                     & $\mQ\wedge\mW$     & $\mV\wedge\mW$             \\[0.5mm]
    $\sV$        & $\mQ\cup\mQ'$       & $\mV\cup\mV'$          & $\left[\mW\cup\mW'\right]_{2}$      & $\mQ\cup\mV$        & $\mQ\cup\mW$       & $\mV\cup\mW$             \\[0.5mm]
    $\aV$        &                     & $\mV\cap\mV'$          & $\left[\mW\cap\mW'\right]_{2}$      & $\mQ\cap\mV$        & $\mQ\cap\mW$       & $\mV\cap\mW$             \\[0.5mm]
    $\sW$        & $\mQ\ast\mQ'$      & $\mV\ast\mV'$         & $\mW\ast\mW'$     & $\mQ\ast\mV$       & $\mQ\ast\mW$      & $\left[\mV\ast\mW\right]_{2}$             \\\noalign{\smallskip}\hline
    \end{tabular}
    \caption{Multiplet products of $S_5$ (see text for explanations).}
    \label{tab:products}
    \end{table}

\subsection{Multiplet products}

It is also useful to work out the irreducible product representations of the group $S_5$.
The possible multiplet products are collected in Table~\ref{tab:products}. %
For example, two quartets produce a singlet, a quartet, a quintet and a sextet:
\begin{equation}
   \sQ \otimes \sQ = \sS \oplus \sQ \oplus \sV \oplus \sW\,.
\end{equation}
The singlet is  the dot product of the two quartet vectors.
For the remaining products we introduced the operator symbols $\times$ (cross), $\vee$ (vee), $\wedge$ (wedge), $\cup$ (cup), $\cap$ (cap) and $\ast$ (star)
to produce antisinglets, quartets, antiquartets, quintets, antiquintets and sextets, respectively.
The subscript `2' in the table implies that there are two different multiplets of the same type, e.g.
\begin{equation}
   \sV \otimes \sW = \sQ \oplus \aQ \oplus \sV \oplus \aV \oplus \sW \oplus \sW\,.
\end{equation}
The products of antimultiplets are identical, e.g.
\begin{equation}
   \aQ \otimes \aQ = \sS \oplus \sQ \oplus \sV \oplus \sW\,.
\end{equation}
For products of multiplets and antimultiplets, the  operations in Table~\ref{tab:products}
are identical but
generate the respective antimultiplet, e.g.
\begin{equation}
   \aV \otimes \sW = \aQ \oplus \sQ \oplus \aV \oplus \sV \oplus \sW \oplus \sW\,,
\end{equation}     
where the $\aQ$ corresponds to $\mV \vee \mW$, etc.
The explicit expressions for the products are given in App.~\ref{sec:products}.

\begin{figure}[t]
   \includegraphics[width=0.4\columnwidth]{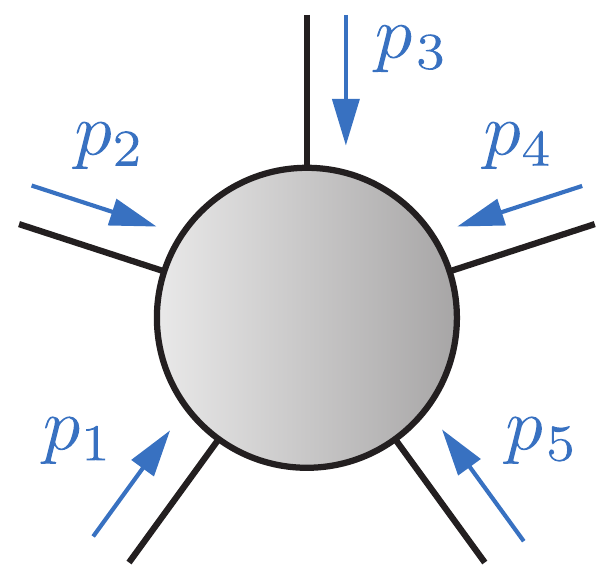} 
   \caption{A generic five-point function.} 
   \label{Fig: Five-point function} 
\end{figure}

\section{Examples}\label{5pa}

\subsection{Five-point functions}\label{5pf}

To illustrate the procedure described above, we consider a five-point correlation function in quantum field theory 
as depicted in Fig.~\ref{Fig: Five-point function}. It depends on five incoming momenta $p_1^\mu \dots p_5^\mu$, 
which in four spacetime dimensions are four-vectors ($\mu=1 \dots 4$ in Euclidean conventions).
The sum $p_1^\mu + \cdots + p_5^\mu = 0$ vanishes because of momentum conservation. 

To arrange these momenta in $S_5$ multiplets, one can  directly apply Eqs.~(\ref{example-1}--\ref{example-2}). 
The singlet vanishes due to momentum conservation,
and the remaining four independent momenta form a quartet:
\begin{equation} \label{jacobi-mom}
   \mQ =  \left[ \begin{array}{c}         p_1^\mu - p_2^\mu \\ 
                                            \frac{1}{\sqrt{3}} \,( p_1^\mu + p_2^\mu -2p_3^\mu) \\
                                            \frac{1}{\sqrt{6}} \,( p_1^\mu + p_2^\mu + p_3^\mu - 3p_4^\mu) \\
                                            \frac{1}{\sqrt{10}}\,( p_1^\mu + p_2^\mu + p_3^\mu + p_4^\mu - 4p_5^\mu ) \end{array}
                                            \right] = \left[ \begin{array}{c}  p^\mu \\ q^\mu \\ k^\mu \\ l^\mu \end{array}\right].
\end{equation}
These are the four independent Jacobi momenta.

Next, we work out the Lorentz invariants of the system.
From four independent momenta one obtains $4+3+2+1=10$ Lorentz invariants, which we denote by
\begin{equation}\label{li-variables-jacobi} \renewcommand{\arraystretch}{1.4}
   \begin{array}{l}
      p^2\,, \\
      q^2\,, \\
      k^2\,, \\
      l^2\,, 
   \end{array}  \qquad
   \begin{array}{rl}
      \omega_1 &\!\!= \frac{2}{\sqrt{3}}\,p\cdot q\,, \\[2mm]
      \omega_2 &\!\!= \sqrt{\frac{2}{3}}\,p\cdot k\,, \\[2mm]
      \omega_3 &\!\!= \sqrt{2}\,q\cdot k\,, 
   \end{array}   \qquad
   \begin{array}{rl}
      \omega_4 &\!\!= \sqrt{\frac{2}{5}}\,p\cdot l\,, \\[2mm]       
      \omega_5 &\!\!= \sqrt{\frac{6}{5}}\,q\cdot l\,, \\[2mm]
      \omega_6 &\!\!= \sqrt{\frac{12}{5}}\,k\cdot l\,.
   \end{array}  
\end{equation} 
The prefactors in the definitions are merely for later convenience.
These variables can  be arranged into $S_5$ multiplets using the  product representations in Table~\ref{tab:products}:
The first column therein shows that from one quartet, Eq.~\eqref{jacobi-mom}, one obtains a singlet $\mQ\cdot \mQ$, a quartet $\mQ \vee\mQ$, a quintet $\mQ \cup \mQ$ 
and a sextet $\mQ \ast \mQ$. However,
the sextet is totally antisymmetric and vanishes if the input quartets are identical (cf.~Table~\ref{tab:4x4} in Appendix~\ref{sec:products}),
so altogether there are ten Lorentz invariants.
The resulting singlet, quartet and quintet obtained via Eq.~\eqref{multiplets-final} are
\begin{align}
   \mQ\cdot \mQ  &= p^2+q^2+k^2+l^2 \,, \label{qveeq0} \\
   \frac{1}{\sqrt{3}}\,\mQ \vee \mQ  &= \left[ \begin{array}{c} \omega_1 + \omega_2 + \omega_4 \\ 
                     \frac{1}{\sqrt{3}}\,(p^2-q^2+\omega_3 + \omega_5) \\ 
                     \frac{1}{\sqrt{6}}\,(p^2 + q^2 - 2k^2  + \omega_6) \\ 
                     \frac{1}{\sqrt{10}}\,(p^2 + q^2 + k^2 - 3l^2) \end{array}\right], \label{qveeq} \\      
       \frac{1}{\sqrt{6}}\,\mQ \cup \mQ  &= \left[ \begin{array}{c} \frac{1}{\sqrt{6}}\,(p^2+q^2 -2k^2 -5 \omega_6) \\ 
                         p^2 -q^2 - 2\omega_3 \\ 
                         \sqrt{3}\,(\omega_1 - 2 \omega_2) \\ 
                         \omega_1 + \omega_2 - 5 \omega_4\\
                         \frac{1}{\sqrt{3}}\,(p^2 - q^2 + \omega_3 - 5 \omega_5)  \end{array}\right] . \label{qveeq1}
\end{align}  

Abbreviating $x_i = p_i^2$, and using the rotation matrices in  Appendix~\ref{sec:li-relations}, 
the singlet and quartet can also be written as
\begin{align}
   \mQ\cdot \mQ  &= 2\,(x_1 + x_2 + x_3 + x_4 + x_5)\,,   \\ 
   \frac{1}{\sqrt{3}}\,\mQ \vee \mQ  &= 2\left[ \begin{array}{c} x_1-x_2 \\ 
                     \frac{1}{\sqrt{3}}\,(x_1+x_2-2x_3) \\ 
                     \frac{1}{\sqrt{6}}\,(x_1+x_2+x_3-3x_4) \\ 
                     \frac{1}{\sqrt{10}}\,(x_1+x_2+x_3+x_4-4x_5) \end{array}\right],  \label{quartet-xi}
\end{align}
while the quintet also depends on the variables $p_i\cdot p_j$. 
If the external particles are onshell and have the same mass  
($x_1=x_2=x_3=x_4=x_5$), the singlet is  a constant and the quartet vanishes,
which leaves only the quintet. These are then the five kinematic variables on which the amplitude depends,
i.e., the analogue of the Mandelstam plane for a four-point function.
Setting $\mQ \vee \mQ= 0$ in Eq.~\eqref{qveeq} fixes the variables $l^2$, $\omega_4$, $\omega_5$ and $\omega_6$,
so that the singlet and quintet turn into
\begin{align}
   \mQ\cdot \mQ  &= \frac{4}{3}\,(p^2+q^2+k^2) \,, \\  
       \frac{1}{\sqrt{6}}\,\mQ \cup \mQ  &= \left[ \begin{array}{c} \sqrt{6}\,(p^2+q^2 -2k^2) \\ 
                         p^2 -q^2 - 2\omega_3 \\ 
                         \sqrt{3}\,(\omega_1 - 2 \omega_2) \\ 
                         6\,(\omega_1 + \omega_2)\\
                        2\sqrt{3}\,(p^2 - q^2 + \omega_3)  \end{array}\right] .
\end{align}  
Thus, the kinematic dependence can be expressed through the variables $p^2$, $q^2$, $k^2$, $\omega_1$, $\omega_2$ and $\omega_3$,
where the sum $p^2 + q^2 + k^2$ is fixed.
If the external particles have different masses (and hence the permutation symmetry is broken), 
the quartet does not vanish but is constant, while the quintet still contains the independent  variables.

This is  analogous to the case of four-point functions, where the
construction based on the permutation group $S_4$ yields six Lorentz invariants forming
a singlet, a doublet and a triplet~\cite{Eichmann:2015nra}. 
For identical onshell particles, the singlet is a constant and the triplet vanishes,
and the two remaining doublet variables  form the Mandelstam plane.

\begin{figure}[t]
   \includegraphics[width=0.9\columnwidth]{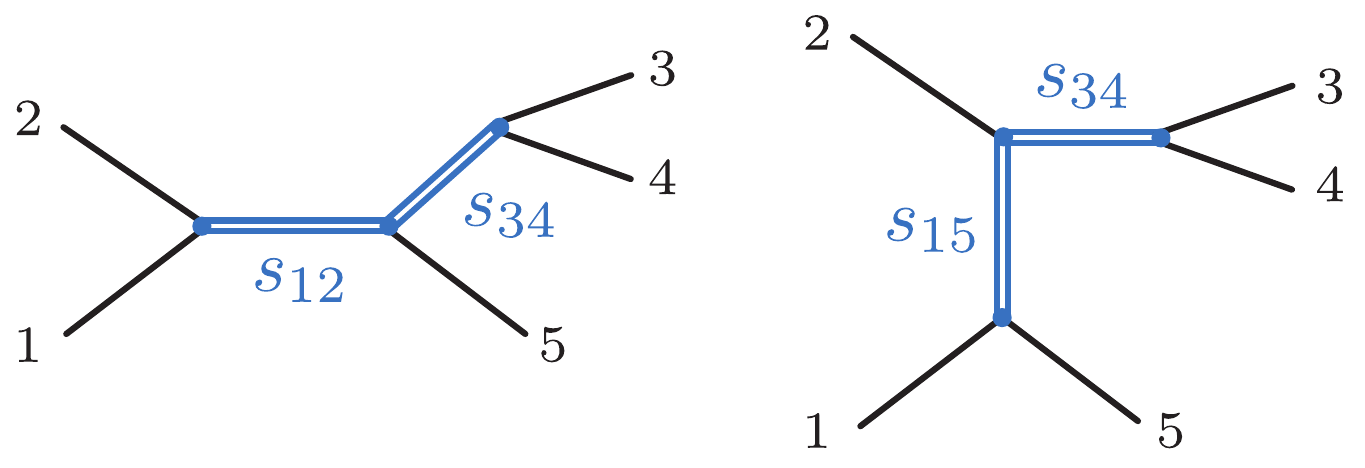} 
   \caption{Exemplary diagrams with intermediate poles.} 
   \label{fig:2-body-top} 
   \vspace{-2mm}
\end{figure}

The Mandelstam variables also allow for a simple discussion of the intermediate poles of the five-point function,
which can describe bound states or resonances. 
There are ten possible ways to form  two- and three-body poles:
\begin{equation}
   \begin{split}
      (12)(345), \quad (13)(245), \quad (14)(235), &\quad (15)(234), \\
                       (23)(145), \quad (24)(135), &\quad (25)(134), \\
                                        (34)(125), &\quad (35)(124),  \\
                                                   &\quad (45)(123).
   \end{split}
\end{equation}
From these one can define ten Mandelstam variables 
\begin{equation}\label{mandelstams}
   s_{ij} = (p_i + p_j)^2,
\end{equation}
e.g. $s_{12} = (p_1+p_2)^2 = (p_3+p_4+p_5)^2$. The intermediate resonances then appear
at fixed values of $s_{ij} = const.$ Fig.~\ref{fig:2-body-top} shows exemplary topologies
in $2\to 3$ scattering processes with intermediate $s$-channel, $t$-channel and isobar poles.
The $s_{ij}$ are linear combinations of the variables in Eq.~\eqref{li-variables-jacobi},
and thus the singlet, quartet and quintet can equally be expressed  in terms of the $s_{ij}$. Abbreviating
\begin{equation}
\begin{split}
   u_1 &= s_{13} + s_{14} + s_{15} - ( s_{23} + s_{24} + s_{25} )\,, \\
   u_2 &= s_{12} + s_{14} + s_{15} - (s_{23} + s_{34} + s_{35}) \,, \\
   u_3 &= s_{12} + s_{13} + s_{15} - (s_{24} + s_{34} + s_{45})\,, \\
   u_4 &= s_{12} + s_{13} + s_{14} - (s_{25} + s_{35} + s_{45})
\end{split}
\end{equation}
and
\begin{equation} \label{vi}
\begin{array}{rl}
   v_1 &= s_{12} - s_{34}\,, \\
   v_2 &= s_{13} - s_{14}\,, 
\end{array}\quad
\begin{array}{rl}
   v_3 &= s_{12} + s_{15} - 2s_{34}\,,  \\
   v_4 &= s_{35} - s_{45} \,, \\
   v_5 &= s_{35} + s_{45} - 2 s_{34}\,,
\end{array}
\end{equation}
they are given by
\begin{align}
   \mQ\cdot \mQ  &= \frac{2}{3}\sum_{i < j} s_{ij} \,,  \nonumber \\
   \frac{1}{\sqrt{3}}\,\mQ \vee \mQ  &= 2\left[ \begin{array}{c} u_1 \\ 
                     \frac{1}{\sqrt{3}}\,(2u_2-u_1) \\ 
                     \frac{1}{\sqrt{6}}\,(3u_3-u_2-u_1) \\ 
                     \frac{1}{\sqrt{10}}\,(4u_4-u_3-u_2-u_1) \end{array}\right], \label{uv-var}  \\      
       \frac{1}{\sqrt{6}}\,\mQ \cup \mQ  &= \left[ \begin{array}{c} 
                     -\sqrt{\frac{2}{3}}\,(u_1-2u_2+3v_1-3v_4) \\ 
                     -2u_1+u_2+u_3+2u_4-6v_1+3v_5 \\ 
                     -\sqrt{3}\,(u_2-u_3+2v_2+v_4) \\ 
                     2u_1-3\,(u_2+u_3-2v_3+v_5) \\
                     -\frac{1}{\sqrt{3}}\,(2u_1-u_2-3u_3+6v_1+3v_4 ) \end{array}\right].  \nonumber
\end{align} 
If the squared momenta $x_i$ ($i=1 \dots 5$) of the external particles are fixed, then the singlet $\mQ\cdot \mQ$ and the quartet variables $u_1 \dots u_4$ are also fixed,
which leaves the five independent quintet variables $v_1 \dots v_5$. If the masses of the particles are the same, then the $x_i$ are identical and the $u_i$ vanish.

\begin{figure}[t]
   \includegraphics[width=0.7\columnwidth]{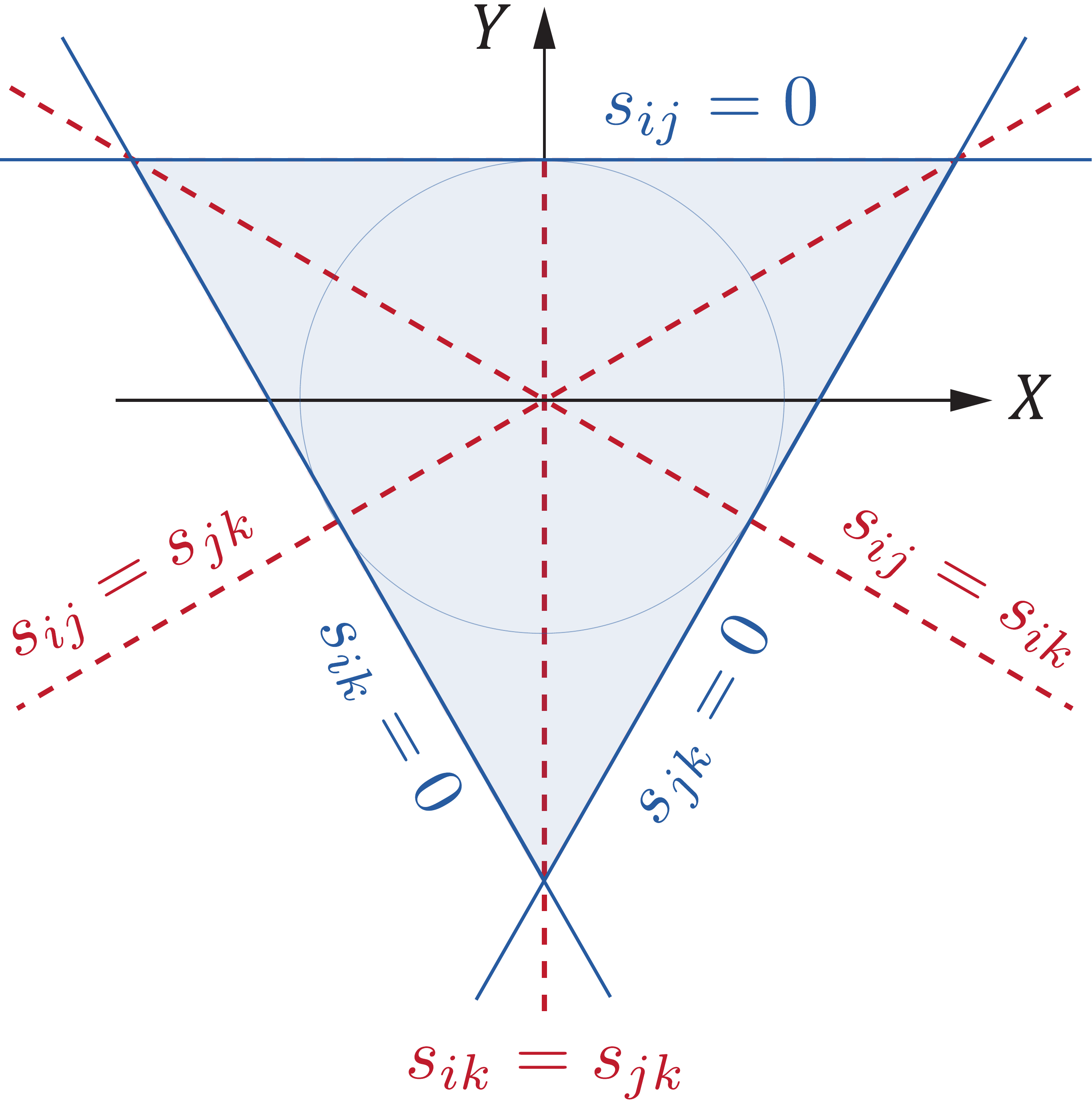} 
   \caption{Mandelstam plane in the variables $s_{ij}$, $s_{ik}$ and $s_{jk}$.} 
   \label{fig:mandelstam} 
\end{figure}

The situation here is more complicated compared to  a four-point function, in which case there are only three Mandelstam variables $s_{12} = s_{34}$,
$s_{13} = s_{24}$ and $s_{14} = s_{23}$ (out of six variables in total). Those only appear in the $S_4$ singlet and doublet but not in the triplet,
so that for a fixed singlet variable the pole conditions $s_{ij} = const.$
correspond to lines in the Mandelstam plane spanned by the doublet variables. 
By contrast, for a five-point function the poles are not restricted to a single multiplet
but distributed over the quartet and quintet.

One can still draw a two-dimensional Mandelstam plane by projecting onto two variables $s_{ik}$ and $s_{jk}$.
This yields the usual Dalitz plot where the poles in $s_{ik}$, $s_{jk}$ and $s_{ij}$ appear as lines.
A symmetric version of this plot is obtained by using the axes
\begin{equation}
    X = \frac{\sqrt{3}\,(s_{ik} - s_{jk})}{s_{ik} + s_{jk} + s_{ij}}\,, \quad
    Y = \frac{s_{ik} + s_{jk} - 2s_{ij}}{s_{ik} + s_{jk} + s_{ij}} 
\end{equation}
as shown in Fig.~\ref{fig:mandelstam}. 
The lines $s_{ik}=0$, $s_{jk}=0$ and $s_{ij}=0$ form a triangle whose corners are
\begin{equation}\renewcommand{\arraystretch}{1.0}
   \left[ \begin{array}{c} 0 \\ -2 \end{array}\right], \quad 
   \left[ \begin{array}{c} \sqrt{3} \\ 1 \end{array}\right], \quad 
   \left[ \begin{array}{c} -\sqrt{3} \\ 1 \end{array}\right]\,.
\end{equation}
For example, for $\{i,j,k\} = \{3,4,5\}$ the variables 
$X=\sqrt{3}\,v_4/v_{345}$ and $Y=v_5/v_{345}$ constructed from $v_4$ and $v_5$ in Eq.~\eqref{vi},
with $v_{345} = s_{35} + s_{45} + s_{34}$, form such a plane with
the poles in $s_{35}$, $s_{45}$ and $s_{35}$  forming a triangle.
Since the $s_{ij}$ are linear combinations of the $u_1 \dots u_4$ and $v_1 \dots v_5$
(see Appendix~\ref{sec:li-relations} for the respective rotation matrix $\mathsf{M}'$),
the same projection can be easily done for any other combination of $\{i,j,k\}$.

\newpage
 
If we allow the $x_i$ to take any spacelike value $\in \mathds{R}_+$, the quartet in Eq.~\eqref{quartet-xi} forms 
the interior of a 
 pentahedron in four dimensions, a so-called 5-cell. To this end, we take the radius
\begin{equation}
   r^2 = \left(\frac{1}{\sqrt{3}}\,\mQ \vee \mQ\right)^2 = \frac{32}{5} \left( \sum_{i=1}^5 x_i^2 - \frac{1}{2}\sum_{i< j} x_i\,x_j \right)
\end{equation}
and consider the unit vector 
\begin{equation}
    \hat\mQ  = \frac{\mQ \vee \mQ}{\sqrt{3}\,r}\,.
\end{equation}
Its five corners $\mC_i$, $i=1 \dots 5$, are defined by the kinematic limits
where $x_i \neq 0$ and the other four $x_{j\neq i}$ vanish:
\begin{align}\label{kin-limits-pentahedron}
\mC_1 &= \frac{1}{4\sqrt{3}}\left[\begin{array}{c} \sqrt{30} \\ \sqrt{10} \\ \sqrt{5} \\ \sqrt{3}  \end{array}\right]\!, \;
\mC_2 = \frac{1}{4\sqrt{3}}\left[\begin{array}{c} -\sqrt{30} \\ \sqrt{10} \\ \sqrt{5} \\ \sqrt{3}  \end{array}\right]\!, \\
\mC_3 &= \frac{1}{4\sqrt{3}}\left[\begin{array}{c} 0 \\ -2\sqrt{10} \\ \sqrt{5} \\ \sqrt{3}  \end{array}\right]\!, \;
\mC_4 = \frac{1}{4}\left[\begin{array}{c} 0 \\ 0 \\ -\sqrt{15} \\ 1  \end{array}\right]\!, \;
\mC_5 = \left[\begin{array}{c} 0 \\ 0 \\ 0 \\ -1  \end{array}\right]\!,  \nonumber
\end{align}
with $\mC_i^2 = 1$.
The soft kinematic limits where only one of the $x_i$ vanishes  define the faces of the pentahedron,
i.e., they are the planes spanned by the points $\mC_{j\neq i}$.

\begin{figure}[t]
   \includegraphics[width=0.8\columnwidth]{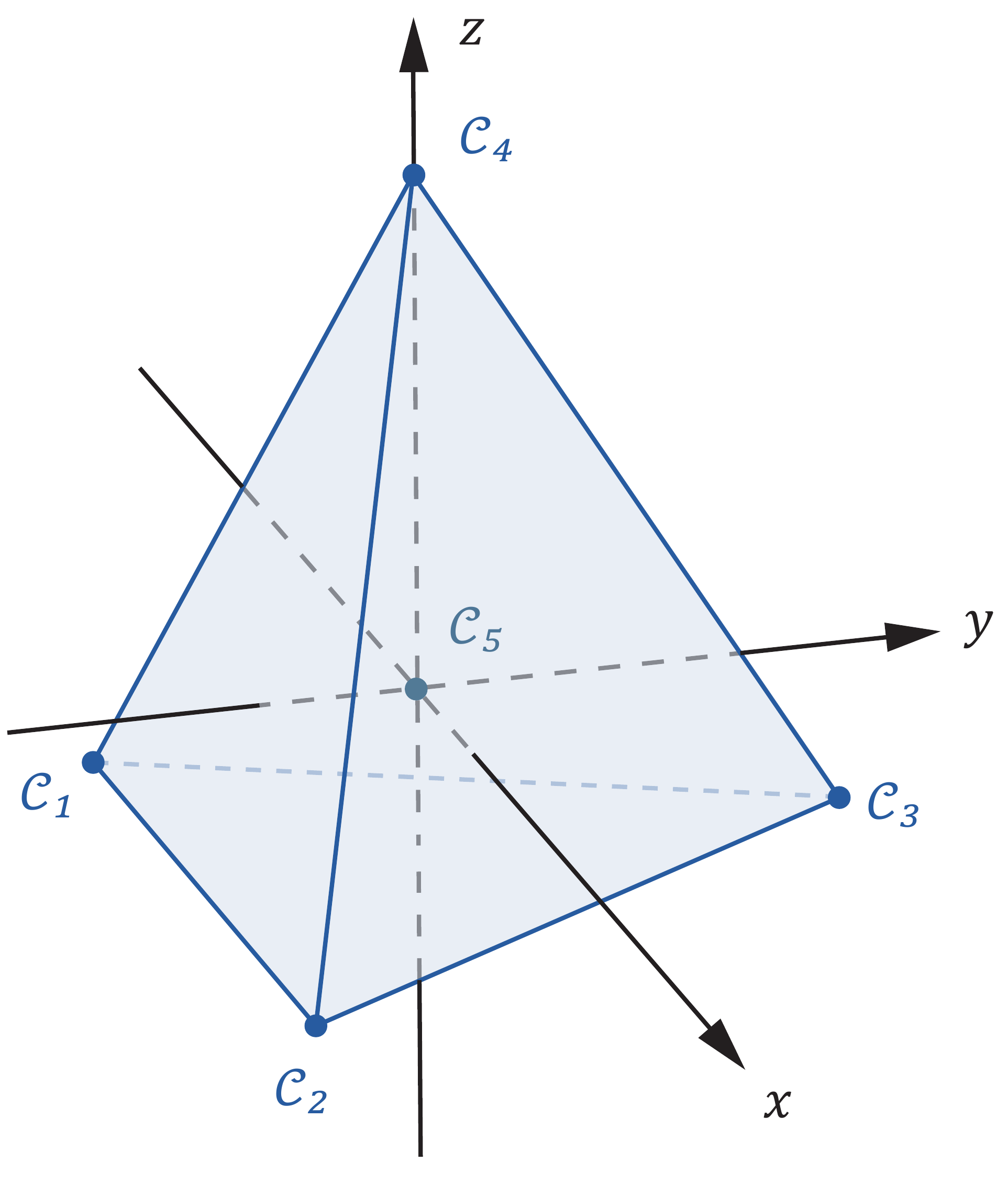} 
   \caption{The stereographic projection of a pentahedron in four dimensions yields a tetrahedron in three dimensions.} 
   \label{fig:tetrahedron} 
   \vspace{5mm}
\end{figure}

This can be visualized through a stereographic projection,
where the unit sphere $S^n$ in $\mathds{R}^{n+1}$ is mapped onto $\mathds{R}^n$. Given a unit vector $(y_0, \dots y_n)$,
the projection is defined by $z_i = y_i/(1-y_0)$ with $i=1 \dots n$.
To this end, we define the projection of the quartet $\hat{\mQ}$ as the three-vector $\widetilde{\mQ}$ with components
\begin{equation}
    \widetilde{\mQ}_i = -3\,\sqrt{\frac{3}{5}}\,\frac{\hat\mQ_i}{1-\hat\mQ_4}\,, \quad i=1\dots 3\,.
\end{equation}
The projections of the $\mC_i$ in Eq.~\eqref{kin-limits-pentahedron} are then given by
\begin{equation}\label{kin-limits-tetrahedron}
\begin{split}
\widetilde\mC_1 = \left[\begin{array}{c} -\sqrt{6} \\ -\sqrt{2} \\ -1  \end{array}\right]\!, \;\;
&\widetilde\mC_2 = \left[\begin{array}{c} \sqrt{6} \\ -\sqrt{2} \\ -1  \end{array}\right]\!, \;\;
\widetilde\mC_3 = \left[\begin{array}{c} 0 \\ 2\sqrt{2} \\ -1  \end{array}\right]\!, \; \\
\widetilde\mC_4 &= \left[\begin{array}{c} 0 \\ 0 \\ 3  \end{array}\right]\!, \;\;
\widetilde\mC_5 = \left[\begin{array}{c} 0 \\ 0 \\ 0   \end{array}\right]  
\end{split}
\end{equation}
and form the tetrahedron shown in Fig.~\ref{fig:tetrahedron}.
Note that the interior of the pentahedron also maps to the exterior of the tetrahedron,
like the projection of the unit circle $S^1$ to $\mathds{R}$ maps half of the circle 
to the interval $[-1,1]$ and the other half to $\mathds{R}\backslash [-1,1]$.
For example, the point $\mC_5$ is mapped to the origin whereas $-\mC_5$ is mapped to $(x,y,z) = ( 0, 0, \infty)$.
If one repeats the procedure, the stereographic projection
of the tetrahedron yields the triangle in Fig.~\ref{fig:mandelstam}, where $\mC_1$, $\mC_2$ and $\mC_3$ are mapped
to the corners of the triangle and $\mC_4$, $\mC_5$ to the origin. 
A similar projection can be done for the quintet in Eq.~\eqref{qveeq1}.

\subsection{Five-gluon vertex}\label{5gv}

As another example, we consider the five-gluon vertex. 
Because it is a five-point function with five Lorentz indices and five color indices,
its general form reads
\begin{equation}\label{5gv-general}
   \Gamma^{\mu\nu\alpha\beta\gamma}(p_1 \dots p_5) = \sum_{ij} F_{ij} (\dots)\,\tau_i^{\mu\nu\alpha\beta\gamma}\,\mathsf{c}^{(j)}_{abcde}\,.
\end{equation}
The $\mathsf{c}^{(j)}_{abcde}$ are the color tensors, and the $\tau_i^{\mu\nu\alpha\beta\gamma}$ 
are the Lorentz tensors which depend on the four independent momenta in the system.
The $F_{ij}(\dots)$ are the scalar dressing functions 
whose arguments are the Lorentz-invariant momentum variables in Eq.~\eqref{li-variables-jacobi}.
Because the vertex must be fully symmetric under exchanging any gluon leg, it is a singlet under $S_5$. 
One can then arrange the dressing functions, Lorentz tensors and color tensors into $S_5$ multiplets such that 
their combination is a symmetric singlet.
Note that the five-gluon vertex does not have a tree-level counterpart in the QCD Lagrangian, so the dressing functions are entirely dynamical.

Let us work out the color factors $\mathsf{c}^{(j)}_{abcde}$ in the multiplet notation. 
They can be expressed in terms of  $f_{abc}$ and $d_{abc}$, the antisymmetric and symmetric structure constants of $SU(N)$, respectively, 
which are given by
\begin{equation}
   \begin{split}
      f_{abc} &=-2i \Tr \left([\mathsf{t}_a,\mathsf{t}_b]\,\mathsf{t}_c \right), \\
      d_{abc} &= 2  \Tr \left(\{\mathsf{t}_a,\mathsf{t}_b\}\,\mathsf{t}_c \right)\,.
   \end{split}
\end{equation}
The  $\mathsf{t}_a$ are the $SU(N)$ generators in the fundamental representation.
We restrict ourselves to  $SU(3)$, where $\mathsf{t}_a=\lambda_a/2$ and $\lambda_a$ are the Gell-Mann matrices. 

There are four types of color structures that can be used as seed elements for obtaining all further multiplets:
\begin{equation} \renewcommand{\arraystretch}{1.3}
    \begin{array}{rl}
      &\delta_{ab}\,f_{cde}\,, \\
      &\delta_{ab}\,d_{cde}\,, 
    \end{array}\qquad
    \begin{array}{rl}
    &f_{abr}\,f_{cds}\,f_{ers}\,, \\
    &f_{abr}\,f_{cds}\,d_{ers}\,.
    \end{array}
\end{equation}
Because the three permutations of $d_{abr}\,d_{cdr}$ can be expressed in terms of the permutations of $f_{abr}\,f_{cdr}$,
and the three permutations of $d_{abr}\,f_{cdr}$ in terms of those of $f_{abr}\,d_{cdr}$ (see e.g. Refs.~\cite{Eichmann:2015nra,Haber:2019sgz}),
the remaining tensors 
\begin{equation}
   d_{abr}\,d_{cds}\,d_{ers}\,, \quad
   f_{abr}\,d_{cds}\,f_{ers}\,, \quad
   d_{abr}\,f_{cds}\,f_{ers}
\end{equation} 
are related to the ones above.
The same is true for combinations like $f_{ars}\, f_{brs} \, f_{cde}$ because of $f_{ars}\,d_{brs} = 0$ and
$\frac{1}{3} \,f_{ars}\,f_{brs} = \frac{3}{5} \,d_{ars}\,d_{brs} = \delta_{ab}$.

We start with the elements $\delta_{ab}\,f_{cde}$ and  $\delta_{ab}\,d_{cde}$.
Each of them has ten possible permutations, which gives 20 linearly independent structures in total.
In the following we replace $\{a,b,c,d,e\} \to \{1,2,3,4,5\}$ and use the shorthand notation
\begin{equation}
\begin{array}{rl}
   f_{ij} &= \delta_{ij}\,f_{klm}\,, \\
   d_{ij} &= \delta_{ij}\,d_{klm}\,, 
\end{array} \qquad  
            k < l < m\,,
\end{equation}
where $\{i,j,k,l,m\}$ is a permutation of $\{1,2,3,4,5\}$,
for example $f_{32} = f_{23} = \delta_{23}\,f_{145} = \delta_{bc}\,f_{ade}$.
Applying the formulas in Eq.~\eqref{multiplets-final},
the ten  permutations of $d_{ij}$ can  be grouped into a singlet, a quartet and a quintet,
\begin{equation}
\begin{split}
   \mS &= \sum_{i<j} d_{ij}\,, \\ 
   \mQ^+ &= \left[ \begin{array}{c} \phi_1-\phi_2 \\ 
                     \frac{1}{\sqrt{3}}\,(\phi_1+\phi_2-2\phi_3) \\ 
                     \frac{1}{\sqrt{6}}\,(\phi_1+\phi_2+\phi_3-3\phi_4) \\ 
                     \frac{1}{\sqrt{10}}\,(\phi_1+x_2+\phi_3+\phi_4-4\phi_5) \end{array}\right], \\
   \mV^+ &= \left[ \begin{array}{c} \sqrt{\frac{2}{3}}\,(\phi_1^- + \phi_2^- - \phi_3^-) \\ 
                  2\phi_1^+ - \phi_2^+ - \phi_3^+ \\ 
                  \sqrt{3}\,(\phi_2^+ - \phi_3^+) \\ 
                  \phi_2^- + \phi_3^- \\
                  \frac{1}{\sqrt{3}}\,(2\phi_1^- - \phi_2^- + \phi_3^-)\end{array}\right],
\end{split}
\end{equation}
where 
\begin{equation}\label{phi_ipm}
\begin{split} 
    \phi_i = \sum_{j \neq i} d_{ij}\,,  \quad\quad
    & \phi_{ij} = d_{ij} - d_{i5} - d_{j5} + \frac{2}{3}\,\phi_5\,, \\
    \phi_1^\pm &= \phi_{12} \pm \phi_{34}\,, \\
    \phi_2^\pm &= \phi_{13} \pm \phi_{24}\,, \\
    \phi_3^\pm &= \phi_{14} \pm \phi_{23}\,.
\end{split} 
\end{equation}

 \pagebreak
The ten permutations of $f_{ij}$ return an antiquartet and a sextet,
\begin{equation}
\begin{split}
   \mQ^- &= \left[ \begin{array}{c} \psi_1-\psi_2 \\ 
                     \frac{1}{\sqrt{3}}\,(\psi_1+\psi_2-2\psi_3) \\ 
                     \frac{1}{\sqrt{6}}\,(\psi_1+\psi_2+\psi_3-3\psi_4) \\ 
                     \frac{1}{\sqrt{10}}\,(\psi_1+x_2+\psi_3+\psi_4-4\psi_5) \end{array}\right] , \\
   \mW &= \left[ \begin{array}{c} 2\left(f_{45} - \frac{1}{5}\,(\psi_4-\psi_5)\right) \\
                     \frac{1}{\sqrt{2}}\left( 3f_{35} - f_{45} + \frac{1}{5}\,(3\psi_3 + \psi_4 - 4\psi_5) \right) \\ 
                     \sqrt{\frac{3}{10}}\left( 4f_{34} - f_{35} - f_{45} - \psi_3 + \psi_4 \right) \\ 
                     \sqrt{\frac{3}{2}}\left( f_{15} + f_{25} + \frac{1}{5}\,(\psi_1 - \psi_2) \right)\\
                     \frac{1}{\sqrt{10}}\,(-2f_{12} + f_{13} + 3f_{14} + f_{23} + 3 f_{24})\\
                     \frac{2}{\sqrt{5}}\,(f_{12} + f_{13} + f_{23}) \end{array}\right], 
\end{split}
\end{equation}  
with
\begin{equation} \renewcommand{\arraystretch}{1.2}
\begin{array}{rl} 
   \psi_1 &= f_{12} - f_{13} + f_{14} - f_{15}\,, \\ 
   \psi_2 &= -f_{21} + f_{23} - f_{24} + f_{25}\,, \\ 
   \psi_3 &= f_{31} - f_{32} + f_{34} - f_{35}\,, \\ 
   \psi_4 &= -f_{41} + f_{42} - f_{43} + f_{45}\,, \\ 
   \psi_5 &= f_{51} - f_{52} + f_{53} - f_{54}\,,
\end{array}\quad \sum_{i=1}^5 \psi_i = 0\,.
\end{equation}

The seed $f_{abm}\,f_{cdn}\,f_{emn}$ in principle has 15 possible permutations,
but it turns out that only six of those are linearly independent.  
Writing $\Psi_{ij,kl} = f_{ijr}\,f_{kls}\,f_{mrs}$, they can be chosen as
\begin{equation}
\begin{array}{rl}
  \Psi_1 &\!\!= \Psi_{12,34}\,, \\
  \Psi_2 &\!\!= \Psi_{13,24}\,, \\ 
  \Psi_3 &\!\!= \Psi_{14,23}\,, 
\end{array} \qquad
\begin{array}{rl}
  \Psi_4 &\!\!= \Psi_{12,35}\,, \\ 
  \Psi_5 &\!\!= \Psi_{13,25}\,, \\ 
  \Psi_6 &\!\!= \Psi_{14,25}\,.
\end{array} 
\end{equation}
The remaining permutations are linear combinations of these $\Psi_i$.
The $\Psi_i$ can be arranged into a sextet:
\begin{equation*}
\begin{split}
   \mW' &= \left[ \begin{array}{c} \Psi_2 + \Psi_3 - \Psi_1 \\
                    \frac{1}{\sqrt{2}}\,(\Psi_2 + \Psi_3 - 2\Psi_1) \\ 
                    \sqrt{\frac{3}{10}}\,(4\Psi_4 + \Psi_2 + \Psi_3 - 2\Psi_1) \\ 
                     \sqrt{\frac{3}{2}}\left(\Psi_2 - \Psi_3 \right)\\
                     \frac{1}{\sqrt{10}}\,(8\Psi_5 - 4\Psi_4 - 3\Psi_3 - 5\Psi_2 + 4\Psi_1)\\
                     \frac{1}{\sqrt{5}}\,(6\Psi_6 - 2\Psi_5 - 2\Psi_4 - 3\Psi_3 - \Psi_2 - \Psi_1) \end{array}\right]. 
\end{split}
\end{equation*}  

The seed $f_{abm}\,f_{cdn}\,d_{emn}$ also has six linearly independent permutations.
Writing $\Phi_{ij,kl} = f_{ijr}\,f_{kls}\,d_{mrs}$ and
\begin{equation}
\begin{array}{rl}
  \Phi_1 &\!\!= \Phi_{12,34}\,, \\
  \Phi_2 &\!\!= \Phi_{13,24}\,, \\ 
  \Phi_3 &\!\!= \Phi_{14,23}\,, 
\end{array} \qquad
\begin{array}{rl}
  \Phi_4 &\!\!= \Phi_{12,35}\,, \\ 
  \Phi_5 &\!\!= \Phi_{13,25}\,, \\ 
  \Phi_6 &\!\!= \Phi_{14,25}\,,
\end{array} 
\end{equation}
its permutations are linear combinations of the $\Phi_i$ and the $d_{ij}$ above.
One obtains an antisinglet and an antiquintet, 
\begin{equation}
\begin{split}
   \mA &= \Phi_1 - \Phi_2 + \Phi_3 , \\
   \mV^- &= \left[ \begin{array}{c} \sqrt{\frac{2}{3}}\,(\Phi_1^- + \Phi_2^- - \Phi_3^-) \\ 
                  2\Phi_1^+ - \Phi_2^+ - \Phi_3^+ \\ 
                  \sqrt{3}\,(\Phi_2^+ - \Phi_3^+) \\ 
                  \Phi_2^- + \Phi_3^- \\
                  \frac{1}{\sqrt{3}}\,(2\Phi_1^- - \Phi_2^- + \Phi_3^-)\end{array}\right],  
\end{split}
\end{equation}  
with
\begin{equation}
\begin{split}
   \Phi_1^+ & = 4\Phi_1 - \phi_2^+ + \phi_3^+\,, \\
   \Phi_2^+ & = -4\Phi_2 + \phi_1^+ - \phi_3^+\,, \\
   \Phi_3^+ & = 4\Phi_3 - \phi_1^+ + \phi_2^+\,, \\
   \Phi_1^- & = -4\,(2\Phi_1-2\Phi_2-\Phi_3) -12\,(\Phi_4-\Phi_5-\Phi_6)\\
            & \quad   -6\,(\phi_1^+ - \phi_2^+ ) - 3\phi_1^-\,, \\ 
   \Phi_2^- & =  -4\,(\Phi_1-3\Phi_2-\Phi_3+\Phi_4-5\Phi_5-3\Phi_6)\\
            & \quad  -4\,(2\phi_1^+ - \phi_2^+ - \phi_3^+) - (4\phi_1^-  + \phi_2^- + 2 \phi_3^-)\,, \\ 
   \Phi_3^- & =  -4\,(\Phi_1+2\Phi_2+\Phi_4+\Phi_5 + 3\Phi_6)\\
            & \quad  +2\,(2\phi_1^+ - \phi_2^+ - \phi_3^+) +2\phi_1^-  + 2\phi_2^- +  \phi_3^- \,,
\end{split}
\end{equation}  
where the $\phi_i^\pm$ are given in Eq.~\eqref{phi_ipm}.

In total, we arrived at 32 linearly independent color structures, which now also have definite symmetries under $S_5$ permutations:
a singlet, an antisinglet, a quartet, an antiquartet, a quintet, an antiquintet, and two sextets.\footnote{The 
 $S_4$ construction for the eight color structures of the four-gluon vertex 
gives a singlet, two doublets and an antitriplet \cite{Eichmann:2015nra}.}
As a cross-check, we projected the one-loop color structures $\text{Tr}\left( \mathsf{t}_a\,\mathsf{t}_b\,\mathsf{t}_c\,\mathsf{t}_d\,\mathsf{t}_e \right)$
and $f_{aij}\,f_{bmi}\,f_{clm}\,f_{dkl}\,f_{ejk}$ of the five-gluon vertex  
onto this 32-dimensional basis 
and  find that they are indeed fully covered.

Concerning the Lorentz tensors $\tau_i^{\mu\nu\alpha\beta\gamma}$ in Eq.~\eqref{5gv-general}, 
from four independent momenta ($p$, $q$, $k$, $l$) one can construct four unit vectors $n_i$, $i=1 \dots 4$ using
Gram-Schmidt orthogonalization. A complete, orthonormal tensor basis is then formed by the combinations
\begin{equation}\label{lorentz-basis}
   n_i^\mu\,n_j^\nu\,n_k^\alpha\,n_l^\beta\,n_m^\gamma\,,
\end{equation}
which yields $4^5 = 1024$ tensors. The structures of the type $\delta^{\mu\nu}\,n_k^\alpha\,n_l^\beta\,n_m^\gamma$
and  $\delta^{\mu\nu}\,\delta^{\alpha\beta}\,n_m^\gamma$ are already included here because
the Kronecker delta can be written as the linear combination
\begin{equation}
   \delta^{\mu\nu} = \sum_{i=1}^4 n_i^\mu\,n_i^\nu\,.
\end{equation}
In the reference frame where the $n_i^\mu$ are the Euclidean unit vectors in $\mathds{R}_4$ this is obviously true,
but since the equation is Lorentz-covariant it holds in any frame.

Because in Landau gauge the gluon propagator is fully transverse, 
only the transverse components of the five-gluon vertex are physically relevant.
When applying transverse projectors, no elements with the same label and index can survive:
$p_1^\mu$ is longitudinal, and so are $p_2^\nu$, $p_3^\alpha$, $p_4^\beta$ and $p_5^\gamma$.
At least for counting purposes, this is equivalent to removing any of the vectors $n_i^\mu$ from Eq.~\eqref{lorentz-basis},
so that the index $i$ runs over $i=1,2,3$ only. This leaves $3^5 = 243$ elements, which is therefore the number of transverse Lorentz tensors.

The basis in Eq.~\eqref{lorentz-basis} does not yet implement the permutation symmetry;
in fact, the orthonormalization is rather counterproductive for this goal and one should go back
to the original momenta $p_1 \dots p_5$ (or $p$, $q$, $k$, $l$).
Given that the analogous situation for the four-gluon vertex is already quite involved,  
we refrain from arranging the Lorentz tensor basis of the five-gluon vertex in $S_5$ multiplets. 
Here one would encounter nontrivial constraints due to the dimensionality of spacetime,
and avoiding kinematic singularities when arranging the basis in transverse and non-transverse tensors is a challenging task.
This is left for future work.

Moreover, one could use the momentum multiplets in Eqs.~(\ref{qveeq0}--\ref{qveeq1})
to construct $1024 \times 32$ Lorentz-color tensors which are fully symmetric,
so that also the respective dressing functions $F_{ij}$ in Eq.~\eqref{5gv-general}
are permutation-group singlets. Once  symmetric, they can only depend on 
symmetric Lorentz invariants like $\mQ\cdot\mQ$, the dot products of the quartet and quintet,  and singlets with higher momentum powers.
In practice, the $F_{ij}$ would then mainly depend on the lowest-dimensional singlet $\mQ\cdot\mQ$,
whereas the angular dependence on the remaining variables is presumably weak or even negligible.
Such a planar degeneracy is well documented in other cases where permutation symmetries play a role,
e.g., for two-photon form factors or the three- and four-gluon vertices~\cite{Eichmann:2014xya,Pinto-Gomez:2022brg,Ferreira:2023fva,Aguilar:2023qqd,Aguilar:2024fen,Eichmann:2017wil,Eichmann:2024glq}.

\subsection{Five-body wave functions}\label{5bwf}

\begin{figure}[t]
   \includegraphics[width=1\columnwidth]{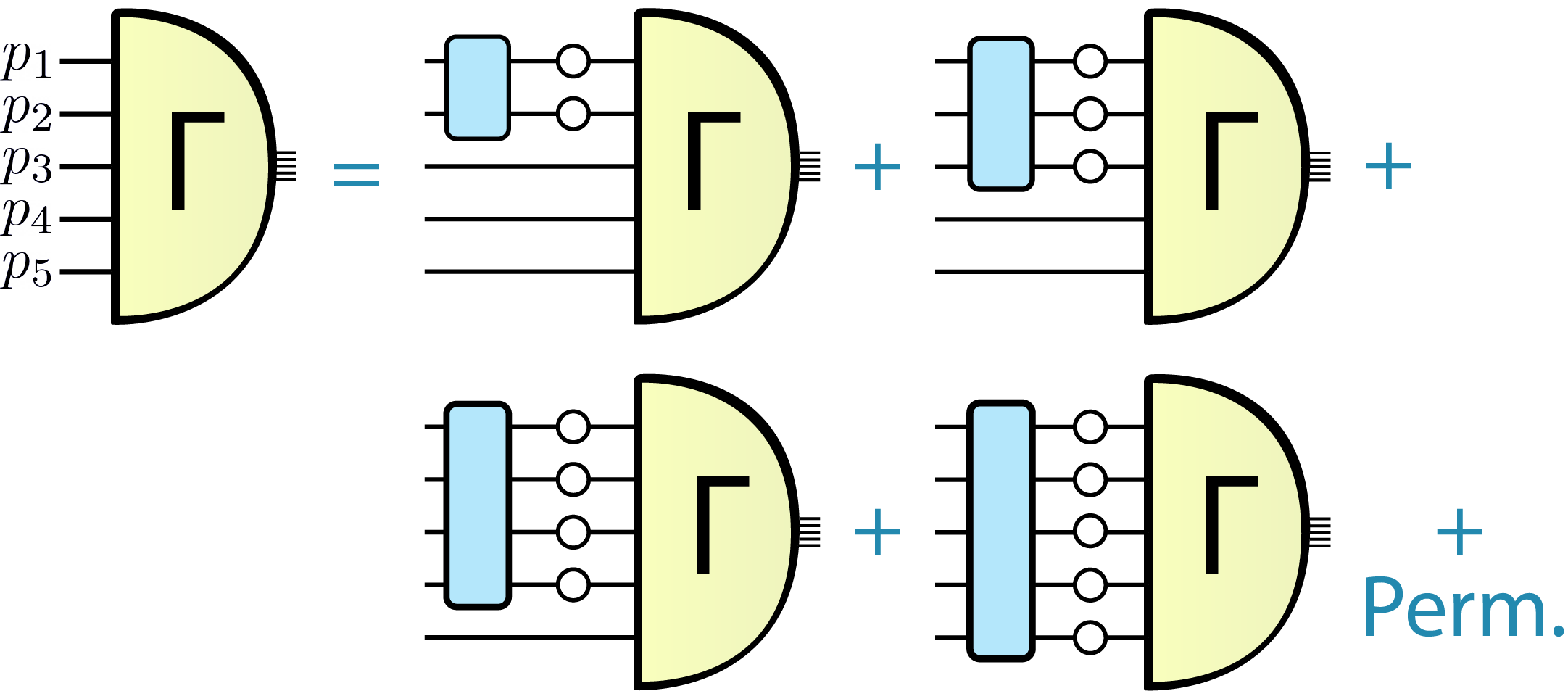} 
   \caption{Generic form of the five-body Bethe-Salpeter equation. The half-circle is the Bethe-Salpeter amplitude,
            the lines with circles are the dressed propagators, and the boxes are the $n$-body kernels.} 
   \label{Fig: Five-Body amplitude} 
\end{figure}

In this section we generalize the discussion in Sec.~\ref{5pf} to five-body wave functions.
The primary application in this context are pentaquarks, whose five-body Bethe-Salpeter equation (BSE)
is shown in Fig.~\ref{Fig: Five-Body amplitude}. The Bethe-Salpeter amplitude,
which is the quantum-field theoretical analogue of a wave function in quantum mechanics,
is represented by the half-circle. It is  the solution of the equation and  encodes the structure of the pentaquark.
The kinematics are the same as in Sec.~\ref{5pf} except that $p_1 + \cdots +  p_5 = P$ is now the total onshell momentum with $P^2 = -m^2$,
where $m$ is the mass of the hadron. 
Thus, the amplitude is actually a six-point function. 
In practice we choose the particle momenta $p_1 \dots p_5$ to be outgoing and the total momentum as incoming,
although this is irrelevant in what follows.

Instead of $p_1 \dots p_5$ it is again convenient to work with four relative momenta $q$, $p$, $k$, $l$, so that the amplitude is written as
$\Gamma(q,p,k,l,P)$. 
Given that a pentaquark is acutally a $qqqq\bar{q}$ state and $p_5$ is the momentum of the antiquark,
instead of using Jacobi momenta 
we redefine these relative momenta as follows:
\begin{equation} \renewcommand{\arraystretch}{2.2}
   \begin{array}{rl}
      q &= \displaystyle\frac{p_1-p_5}{2}\,, \\
      p &= \displaystyle\frac{p_2-p_5}{2}\,, 
   \end{array}\quad
   \begin{array}{rl}
      k &= \displaystyle\frac{p_3-p_5}{2}\,, \\
      l &= \displaystyle\frac{p_4-p_5}{2}\,, 
   \end{array}\quad
      P = \sum_{i=1}^{5} p_i\,.
\end{equation} 
From these  one can  form  15 Lorentz invariants, which we  redefine as 
\begin{equation}\label{li-variables} \renewcommand{\arraystretch}{1.4}
   \begin{array}{l}
      q^2\,, \\
      p^2\,, \\
      k^2\,, \\
      l^2\,, \\        
      P^2\,, 
   \end{array}  \qquad
   \begin{array}{rl}
      \omega_1 &\!\!= q\cdot p\,, \\
      \omega_2 &\!\!= q\cdot k\,, \\
      \omega_3 &\!\!= q\cdot l\,,  \\
      \omega_4 &\!\!= p\cdot k\,, \\        
      \omega_5 &\!\!= p\cdot l\,, \\
      \omega_6 &\!\!= k\cdot l\,, 
   \end{array}  \qquad
   \begin{array}{rl}
      \eta_1 &\!\!= q\cdot P\,, \\
      \eta_2 &\!\!= p\cdot P\,, \\
      \eta_3 &\!\!= k\cdot P\,, \\
      \eta_4 &\!\!= l\cdot P\,.
   \end{array}  
\end{equation}   
Alternatively, one could work with the 15 variables $x_i = p_i^2$ and $x_{ij} = p_i\cdot p_j$.
 
These variables can again be arranged into $S_5$ multiplets.
The additional variables $P^2$ and $\eta_1 \dots \eta_4$ form a singlet and a quartet,
so that one arrives at two singlets, two quartets and one quintet in total.
One singlet is $P^2$ and the other is the symmetric variable
\begin{equation}\label{5b-singlet}
   \mS_0 = \frac{1}{5}\left(  q^2 + p^2 + k^2 + l^2 - \frac{1}{2}\sum_{i=1}^6 \omega_i \right).
\end{equation}
The two quartets are given by
\begin{align} \renewcommand{\arraystretch}{1.2}
   \mQ_1 &= \left[ \begin{array}{c} q^2 - p^2 -\frac{2}{3}\,v_4 \\ 
                  \frac{1}{\sqrt{3}}\,(q^2+p^2-2k^2) -\frac{2}{3}\,v_5   \\ 
                  \frac{1}{\sqrt{6}}\,(q^2+p^2+k^2-3l^2)  - \frac{2}{3}\,v_1  \\ 
                  -\frac{1}{\sqrt{10}} \frac{5}{3}\,(q^2+p^2+k^2+l^2-8\mS_0) \end{array}\right], \label{5b-quartet-1} \\
   \mQ_2 &= \left[ \begin{array}{c} \eta_1 - \eta_2 \\ 
                  \frac{1}{\sqrt{3}}\,(\eta_1+\eta_2-2\eta_3) \\ 
                  \frac{1}{\sqrt{6}}\,(\eta_1+\eta_2+\eta_3-3\eta_4)  \\ 
                  \frac{1}{\sqrt{10}}\,(\eta_1+\eta_2+\eta_3+\eta_4) \end{array}\right],  \label{5b-quartet-2}
\end{align}     
where the $v_i$ are the components of the quintet:
\begin{align} \renewcommand{\arraystretch}{1.2}
   \mV = \left[ \begin{array}{c} v_1 \\ v_2 \\ v_3 \\ v_4 \\ v_5 \end{array}\right] &=
         \left[ \begin{array}{c} \sqrt{\frac{2}{3}}\,(\omega_1 + \omega_2 - \omega_3 + \omega_4 - \omega_5 - \omega_6) \\ 
                  2\omega_1 - \omega_2 - \omega_3 - \omega_4 - \omega_5 + 2\omega_6 \\ 
                  \sqrt{3}\,(\omega_2 - \omega_3 - \omega_4 + \omega_5) \\ 
                  \omega_2 + \omega_3 - \omega_4 - \omega_5 \\
                  \frac{1}{\sqrt{3}}\,(2\omega_1 - \omega_2 + \omega_3 - \omega_4 + \omega_5 - 2\omega_6)\end{array}\right].   \nonumber
\end{align}

   \begin{table}
  \centering
  \begin{tabular}{c @{\quad}  c @{\;\;} c @{\;\;} c @{\;\;} c @{\;\;} c @{\quad} c @{\quad} c } \hline\noalign{\smallskip}
    $n$   & $\mS_0$ & $P^2$ & $\eta_i$ & $p_1^2 \dots p_n^2$ & $\mM$  & Total         & Indep.    \\ \noalign{\smallskip}\hline\noalign{\smallskip}
    $2$        & 1       & 1     & 1        & $-$           & $-$     & 3             & 3         \\[0.5mm]
    $3$        & 1       & 1     & 2        & 2             & $-$     & 6             & 6         \\[0.5mm]
    $4$        & 1       & 1     & 3        & 3             & $2$     & 10            & 10        \\[0.5mm]
    $5$        & 1       & 1     & 4        & 4             & $5$     & 15            & 14        \\[0.5mm]
    $6$        & 1       & 1     & 5        & 5             & $9$     & 21            & 18        \\ \noalign{\smallskip}\hline\noalign{\smallskip}
    $n$        & 1       & 1     & $n-1$    & $n-1$         & $n(n-3)/2$ & $n(n+1)/2$    & $4n-6$    \\\noalign{\smallskip}\hline
  \end{tabular}
  \caption{$S_n$ multiplet counting for  $n$-body systems. Each column stands for a  multiplet with a given number of variables.
           The total number of possible Lorentz invariants is $n(n+1)/2$, of which
           in four spacetime dimensions $4n-6$ are independent.}
  \label{tab:variables}
\end{table}

The multiplet construction can be extended to general $n$-body systems subject to the permutation group $S_n$.
A two-body system forms two singlets ($q^2$ and $P^2$) and an antisinglet ($q\cdot P$).
A three-body system gives two singlets ($\mS_0$ and $P^2$) and two doublets~\cite{Eichmann:2011vu}.  
A four-body system yields two singlets ($\mS_0$ and $P^2$), a doublet and two triplets~\cite{Eichmann:2015cra}.
The general case of an $n$-body system is shown in Table~\ref{tab:variables}:

\smallskip
{\tiny$\blacksquare$}
$P^2$ is always a singlet, and for any $n$ one can construct a singlet as the sum $p_1^2 + \dots + p_n^2$.
In our case this is the linear combination $16\mS_0 + P^2/5$ with $\mS_0$ from Eq.~\eqref{5b-singlet}.

\smallskip
{\tiny$\blacksquare$}
A general $n$-body system has $n-1$ relative momenta  and hence $n-1$ angular variables $\eta_i$, which form an $(n-1)$-dimensional multiplet of $S_n$.
In two-, three- and four-body systems these variables have the least impact on the dynamics, and usually they can be safely neglected when solving the $n$-body BSE.
In the five-body case, this is the quartet~\eqref{5b-quartet-2}.

\smallskip
{\tiny$\blacksquare$}
The variables $p_1^2 \dots p_n^2$ form another $(n-1)$-plet, with their sum being constrained by $\mS_0$.
In our case, this would be the linear combination $\frac{8}{5}\,(3\mQ_1 + \mQ_2)$ of the two quartets~(\ref{5b-quartet-1}--\ref{5b-quartet-2}), 
which is identical to the expression in Eq.~\eqref{quartet-xi}.

\smallskip
{\tiny$\blacksquare$}
In total there are $n(n+1)/2$ possible Lorentz invariants, so the difference $n(n-3)/2$ gives another multiplet $\mM$. 
In a four-body system, this is a doublet which contains the intermediate two-body poles and
introduces a resonance mechanism in the BSE. Because pions are the lightest two-body clusters in a four-quark system, 
this is also the most important multiplet next to $\mS_0$~\cite{Eichmann:2015cra}.
In the five-body system it is the quintet $\mV$, but since the intermediate poles are spread across the quintet and quartet
(see Sec.~\ref{5pf}) 
the interpretation is less clear.

The situation is further obscured by another aspect:
For five-body wave functions (which are six-point functions) one encounters for the first time a dimensional constraint relating the Lorentz invariants, 
because $n$ four-vectors can only depend on $4n-6$ independent variables.
This is obvious when 
casting the five vectors in a specific reference frame, schematically
\begin{equation*} \renewcommand{\arraystretch}{1.0}
   P = \left[ \begin{array}{c} 0 \\ 0 \\ 0 \\ \bullet \end{array}\right]\!, \;
   q = \left[ \begin{array}{c} 0 \\ 0 \\ \bullet \\ \bullet \end{array}\right]\!, \;
   p = \left[ \begin{array}{c} 0 \\ \bullet \\ \bullet \\ \bullet \end{array}\right]\!, \;
   k = \left[ \begin{array}{c} \bullet \\ \bullet \\ \bullet \\ \bullet \end{array}\right]\!, \;
   l = \left[ \begin{array}{c} \bullet \\ \bullet \\ \bullet \\ \bullet \end{array}\right]\!, 
\end{equation*}
where the symbol $\bullet$ stands for a nonzero entry.
The five vectors can only have 14 independent entries and thus there can also  only be 14 independent Lorentz invariants,
so not all variables in Eq.~\eqref{li-variables} can be independent.
For a general $n$-body system with $n \geq 5$ independent four-momenta, this gives $4n-6$ independent Lorentz invariants
as opposed to the naive result $n(n+1)/2$. 

In our  case with $n=5$, 
this yields one constraint equation.
Unfortunately, the constraint does not eliminate a single component in a multiplet (which would break the permutation symmetry)
or fix the length of a given multiplet,
but instead relates all variables in Eq.~\eqref{li-variables} through a complicated quintic polynomial.
We derive it in Appendix~\ref{sec:constraint} and state the result  in Eq.~\eqref{dim-constraint}.

In particular, the constraint implies that a symmetric limit of the form $\mV = \mQ_1 = \mQ_2 = 0$, 
where only $\mS_0$ remains as an independent variable, does not exist. 
For example, after setting all $\omega_i=0$ and $p^2=q^2=k^2=l^2$,
the constraint turns into
\begin{equation}
   \sum_{i=j}^4 \eta_j^2 = p^2\,P^2\,.
\end{equation}
Because the total momentum is onshell and $P^2 = -m^2$ is fixed, and because in that case $\mS_0 = \frac{4}{5}\,p^2$,
one can no longer set all $\eta_i=0$ simultaneously.
Alternatively, after setting all $\omega_i=0$ and $\eta_j=0$, the constraint equation turns into 
\begin{equation} 
   q^2\,p^2\,k^2\,l^2\,P^2=0\,,
\end{equation}
so one can no longer choose $p^2=q^2=k^2=l^2$.
Note that in both cases the fourth component of the quartet $\mQ_1$ remains nonzero.
This has practical consequences for the BSE solution: For two-, three- and four-body
systems, it is possible to obtain approximate solutions by eliminating all variables except $\mS_0$. 
In the five-body case, one must keep at least another variable in the 
system.\footnote{A similar constraint arises already for four-point functions, but in that case  it does not eliminate 
a variable and only restricts the possible kinematic regions: 
The triplet  variables do not probe the full tetrahedron but only a subset of it~\cite{Eichmann:2015nra}.}

The dimensional constraint also has consequences for the tensor basis of a pentaquark Bethe-Salpeter amplitude.
Its general form is analogous to Eq.~\eqref{5gv-general},
\begin{equation}
   \Gamma(q,p,k,l,P) = \sum_{i} f_i(\dots)\,\tau_i(q,p,k,l,P)\,,
\end{equation}
where the $\tau_i$ are combinations of Dirac, color and flavor tensors.
The Dirac parts carry six Dirac indices (five for the quarks and antiquarks and one for the pentaquark),
so one must find all linearly independent Dirac tensors with six indices that can be constructed from five momenta.
To obtain an orthonormal basis, one may employ a Gram-Schmidt orthogonalization like the one discussed in connection with Eq.~\eqref{lorentz-basis}.
However, this can only produce four independent unit vectors, so the fifth momentum does not even enter in the basis construction.
While this does reduce the number of tensors, it also makes it more cumbersome to establish manifest $S_5$ symmetry.
Also here a similar constraint  already arises in four-point functions; although it does not remove a momentum entirely,
it reduces the number of basis elements, e.g.,  like in the photon four-point function~\cite{Eichmann:2015nra}.

\section{Summary}\label{summary}

In this paper we discussed applications of the permutation group $S_5$ for five-point functions and five-body wave functions.
Our main result in Eq.~\eqref{multiplets-final} provides a simple algorithmic tool to cast the permutations of
any object of the form $f_{ijklm}$ into $S_5$ multiplets with a definite symmetry under permutations.
Given that studies of higher $n$-point functions are becoming increasingly feasible and interesting for different communities,
our work has various potential applications of which we discussed  a few: the kinematic regions in five-point functions,
the color structure of the five-gluon vertex, and the properties of five-body wave functions.

Our discussion also shows that higher $n$-point functions show nontrivial new effects, 
like the redundancy of Lorentz invariants and tensor basis elements due to dimensional constraints.
These must be worked out in detail and may complicate the study of higher $n$-body systems substantially.
Our present work lays out the groundwork for such studies.

\vspace*{3mm}
{\bf Acknowledgments}\\
We are pleased to acknowledge interactions with  Christian Fischer, Maxwell Hansen, Joshua Hoffer, Maxim Mai and Teresa Peña.
This work was supported by the Portuguese Science fund FCT 
under grant numbers CERN/FIS-PAR/0023/2021 and PRT/BD/152265/2021 
and by the Austrian Science Fund FWF under grant number 10.55776/PAT2089624.
This work contributes to the aims of the USDOE ExoHad Topical Collaboration, contract DE-SC0023598.

\appendix

\newpage

\clearpage

\section{Multiplet coefficients}\label{App:multiplets}

In this appendix we collect the coefficients $a^{(1 \dots 4)}_{nm}$, $b^{(1 \dots 5)}_{nm}$, $c^{(1 \dots 6)}_{nm}$, $d^{(1 \dots 6)}_{m}$ and $e^{(1 \dots 6)}_{m}$
which appear in the expressions for the multiplets in Eq.~\eqref{multiplets-final}.
Using a matrix notation, the $a^{(1 \dots 4)}_{nm}$ entering in the quartets read as follows:
\begin{equation} \renewcommand{\arraystretch}{1.0}
\begin{split}
   a^{(1)} &= \left( \begin{array}{rrrrrr} 1 & 0 & 1 & -1 & 0 & 0 \\ 
                                          0 & 0 & -1 & 0 & 0 & -1 \\
                                          -1 & 0 & 0 & 1 & 0 & 1 \\
                                          0 & -1 & 0 & 0 & 1 & 0 \\
                                          0 & 1 & 0 & 0 & -1 & 0 \end{array}\right), \\
   a^{(2)} &= -\frac{1}{\sqrt{3}}\left( \begin{array}{rrrrrr} 1 & 0 & 1 & 1 & -2 & 0 \\ 
                                          -2 & 0 & 1 & -2 & 0 & 1 \\
                                           1 & -2 & 0 & 1 & 0 & 1 \\
                                          0 & 1 & 0 & 0 & 1 & -2 \\
                                          0 & 1 & -2 & 0 & 1 & 0 \end{array}\right), \\
   a^{(3)} &= -\frac{1}{\sqrt{6}}\left( \begin{array}{rrrrrr} 1 & -3 & 1 & 1 & 1 & 0 \\ 
                                           1 & 0 & 1 & 1 & 0 & 1 \\
                                           1 & 1 & 0 & 1 & -3 & 1 \\
                                          -3 & 1 & -3 & 0 & 1 & 1 \\
                                          0 & 1 & 1 & -3 & 1 & -3 \end{array}\right), \\
   a^{(4)} &= -\frac{1}{\sqrt{10}}\left( \begin{array}{rrrrrr} 1 & 1 & 1 & 1 & 1 & -4 \\ 
                                           1 & -4 & 1 & 1 & -4 & 1 \\
                                           1 & 1 & -4 & 1 & 1 & 1 \\
                                           1 & 1 &  1 & -4 & 1 & 1 \\
                                          -4 & 1 & 1 & 1 & 1 & 1 \end{array}\right).
\end{split}
\end{equation}  
The $b^{(1 \dots 5)}_{nm}$ appearing in the quintets are given by
\begin{equation} \renewcommand{\arraystretch}{1.0}
\begin{split}                                        
   b^{(1)} &= \left( \begin{array}{rrrrrr} 1 & 0 & -2 & 1 & 1 & 0 \\ 
                                           -2 & 0 & 1 & -2 & 0 & 1 \\
                                           1 & 1 & 0 & 1 & 0 & -2 \\
                                           0 & -2 &  0 & 0 & 1 & 1 \\
                                           0 & 1 & 1 & 0 & -2 & 0 \end{array}\right), \\
   b^{(2)} &= \frac{1}{2}\sqrt{\frac{3}{2}}\left( \begin{array}{rrrrrr} 1 & 1 & 1 & -2 & 1 & 2 \\ 
                                           -2 & -4 & -2 & -2 & -4 & -2 \\
                                           -2 & 1 & 2 & 1 & 1 & 1 \\
                                           1 & 1 &  -2 & 2 & 1 & 1 \\
                                           2 & 1 & 1 & 1 & 1 & -2 \end{array}\right), \\
   b^{(3)} &= \frac{3}{2\sqrt{2}}\left( \begin{array}{rrrrrr} 1 & -1 & -1 & 0 & -1 & 2 \\ 
                                           0 & 0 & 0 & 0 & 0 & 0 \\
                                           0 & -1 & 2 & 1 & -1 & -1 \\
                                           1 & 1 &  0 & -2 & 1 & -1 \\
                                           -2 & 1 & -1 & 1 & 1 & 0 \end{array}\right), \\
   b^{(4)} &= \frac{1}{2}\sqrt{\frac{3}{2}}\left( \begin{array}{rrrrrr} -1 & -3 & -1 & -2 & 3 & 0 \\ 
                                           0 & 0 & -2 & 0 & 0 & -2 \\
                                           -2 & 3 & 0 & -1 & -3 & -1 \\
                                           3 & 1 &  0 & 0 & -1 & 3 \\
                                           0 & -1 & 3 & 3 & 1 & 0 \end{array}\right), \\
   b^{(5)} &= \frac{1}{2\sqrt{2}}\left( \begin{array}{rrrrrr} -5 & 3 & 1 & 4 & 1 & 0 \\ 
                                           -2 & 0 & 4 & -2 & 0 & 4 \\
                                           4 & 1 & 0 & -5 & 3 & 1 \\
                                           3 & 1 &  -6 & 0 & -5 & 1 \\
                                           0 & -5 & 1 & 3 & 1 & -6 \end{array}\right),
\end{split}
\end{equation}
and the $c^{(1 \dots 6)}_{nm}$ for the sextets read
\begin{equation} \renewcommand{\arraystretch}{1.0}
\begin{split}                                        
   c^{(1)} &= \left( \begin{array}{rrrrrr} 1 & -2 & 0 & -1 & 1 & 2 \\ 
                                           0 & 2 & 1 & 0 & 2 & 1 \\
                                           -1 & 1 & 2 & 1 & -2 & 0 \\
                                           -2 & 0 &  -2 & 2 & -1 & -1 \\
                                           2 & -1 & -1 & -2 & 0 & -2 \end{array}\right), \\
   c^{(2)} &= -\frac{1}{2\sqrt{2}}\left( \begin{array}{rrrrrr} 1 & 1 & -3 & -4 & -5 & -4 \\ 
                                           6 & 8 & 4 & 6 & 8 & 4 \\
                                           -4 & -5 & -4 & 1 & 1 & -3 \\
                                           1 & -3 &  -2 & -4 & -1 & 5 \\
                                           -4 & -1 & 5 & 1 & -3 & -2 \end{array}\right), \\
   c^{(3)} &= -\frac{1}{2}\sqrt{\frac{15}{2}}\left( \begin{array}{rrrrrr} 1 & 1 & 1 & 0 & -1 & 0 \\ 
                                           -2 & 0 & 0 & -2 & 0 & 0 \\
                                           0 & -1 & 0 & 1 & 1 & 1 \\
                                           1 & 1 &  -2 & 0 & -1 & 1 \\
                                           0 & -1 & 1 & 1 & 1 & -2 \end{array}\right), \\
   c^{(4)} &= \frac{1}{2}\sqrt{\frac{3}{2}}\left( \begin{array}{rrrrrr} -3 & -1 & 3 & -2 & 1 & -4 \\ 
                                           0 & 0 & 2 & 0 & 0 & 2 \\
                                           -2 & 1 & -4 & -3 & -1 & 3 \\
                                           1 & -3 &  0 & 4 & 3 & -1 \\
                                           4 & 3 & -1 & 1 & -3 & 0 \end{array}\right), \\
   c^{(5)} &= \frac{1}{2}\sqrt{\frac{5}{2}}\left( \begin{array}{rrrrrr} -1 & -3 & -3 & -2 & -1 & 0 \\ 
                                           0 & 0 & 2 & 0 & 0 & 2 \\
                                           -2 & -1 & 0 & -1 & -3 & -3 \\
                                           3 & 3 &  0 & 0 & 1 & 1 \\
                                           0 & 1 & 1 & 3 & 3 & 0 \end{array}\right), \\
   c^{(6)} &= \sqrt{5}\left( \begin{array}{rrrrrr} -1 & 0 & 0 & 1 & -1 & 0 \\ 
                                           0 & 0 & -1 & 0 & 0 & -1 \\
                                           1 & -1 & 0 & -1 & 0 & 0 \\
                                           0 & 0 &  0 & 0 & 1 & 1 \\
                                           0 & 1 & 1 & 0 & 0 & 0 \end{array}\right). 
\end{split}
\end{equation}
The coefficients $d^{(1 \dots 6)}_{m}$ and $e^{(1 \dots 6)}_{m}$ for the remaining sextet are given by
\begin{equation} \renewcommand{\arraystretch}{1.0}
\begin{split}
    d^{(1)} &= \left( \begin{array}{rrrrrr} 
                1 & 1 & -1 & -1 & -1 & 1  \end{array}\right), \\
    d^{(2)} &= -\tfrac{1}{2\sqrt{2}}\left( \begin{array}{rrrrrr} 
                1 & -2 & -1 & -1 & 2 & 1  \end{array}\right), \\
    d^{(3)} &= -\tfrac{1}{2}\sqrt{\tfrac{3}{10}}\left( \begin{array}{rrrrrr} 
                1 &  2 & 3 & -1 & -2 & -3  \end{array}\right), \\
    d^{(4)} &= \tfrac{1}{2}\sqrt{\tfrac{3}{2}}\left( \begin{array}{rrrrrr} 
                -1 & 0 & -1 & 1 & 0 & 1  \end{array}\right), \\
    d^{(5)} &= \tfrac{1}{2\sqrt{10}}\left( \begin{array}{rrrrrr} 
                5 & -4 & 1 & -5 & 4 & -1  \end{array}\right), \\
    d^{(6)} &= \tfrac{2}{\sqrt{5}}\left( \begin{array}{rrrrrr} 
               1 & 1 & -1 & -1 & -1 & 1  \end{array}\right), \\
    e^{(1)} &= \left( \begin{array}{rrrrrr}   
               0 & 0 & 0 & 0 & 0 & 0  \end{array}\right), \\
    e^{(2)} &= -\tfrac{3}{2\sqrt{2}}\left( \begin{array}{rrrrrr} 
               1 & 0 & 1 & -1 & 0 & -1  \end{array}\right), \\
    e^{(3)} &= -\tfrac{1}{2}\sqrt{\tfrac{3}{10}}\left( \begin{array}{rrrrrr} 
               3 & -4 & -1 & -3 & 4 & 1  \end{array}\right), \\
    e^{(4)} &= \tfrac{1}{2}\sqrt{\tfrac{3}{2}}\left( \begin{array}{rrrrrr} 
               1 & -2 & -1 & -1 & 2 & 1  \end{array}\right), \\
    e^{(5)} &= \tfrac{1}{2\sqrt{10}}\left( \begin{array}{rrrrrr} 
               -5 & -2 & -7 & 5 & 2 & 7  \end{array}\right), \\
    e^{(6)} &= \tfrac{1}{\sqrt{5}}\left( \begin{array}{rrrrrr} 
               1 & 1 & -1 & -1 & -1 & 1  \end{array}\right).
\end{split}
\end{equation}

\section{Multiplet products}\label{sec:products}

In this appendix we provide details on the multiplet products in Table~\ref{tab:products}.
Their construction follows from the transformation laws in Eq.~\eqref{permutation-tf-s4}.
Consider a given multiplet $\mN$, which is the product of two multiplets $\mM$ and $\mM$'.
We write the transpositions and cyclic permutations generically as
\begin{equation}
  \begin{array}{rll}
  (\pT \,\mN)_m &= A_{mn}\,\mN_n\,, \\ 
  (\pC \,\mN)_m &= B_{mn}\,\mN_n\,, \quad\qquad & m,n = 1 \dots \text{dim}\,\mN\,, \\[2mm]
  (\pT \,\mM)_i &= a_{ij}\,\mM_j\,, \\
  (\pC \,\mM)_i &= b_{ij}\,\mM_j\, \quad & i,j = 1 \dots \text{dim}\,\mM\,, \\[2mm] 
  (\pT \,\mM')_k &= a'_{kl}\,\mM'_l\,, \\
  (\pC \,\mM')_k &= b'_{kl}\,\mM'_l\, \quad & k,l = 1 \dots \text{dim}\,\mM'\,,
  \end{array}
\end{equation}
with matrix representations $A$, $a$, $a'$ and $B$, $b$, $b'$.
Each component of $\mN$ must be  a bilinear combination
\begin{equation}
   \mN_m = c^m_{ik}\,\mM_i \,\mM'_k  \,,
\end{equation}
whose coefficients $c^m_{ik}$ we are after. For a transposition $\pT$, the transformation law  entails
\begin{equation}
\begin{split}
   (\pT \,\mN)_m &= A_{mn}\,c^n_{ik}\,\mM_i \,\mM'_k \\
                 &= c^m_{ik}\,(\pT\,\mM)_i \,(\pT\,\mM')_k   \\
                 &= c^m_{ik}\,a_{ij}\,a'_{kl}\,\mM_j\,\mM_l'\,.
\end{split}
\end{equation}
Comparing the coefficients of $\mM_j\,\mM_l'$, this yields the equations
\begin{equation}
\begin{split}
   A_{mn}\,c^n_{jl} &= a_{ij}\,a'_{kl}\,c^m_{ik}\,,  \\
   B_{mn}\,c^n_{jl} &= b_{ij}\,b'_{kl}\,c^m_{ik} \,.
\end{split}
\end{equation}
In total this gives $2\,(\text{dim}\,\mN)(\text{dim}\,\mM)(\text{dim}\,\mM')$
equations for the $c^m_{ik}$ which must be solved.
For example, the largest system of equations in our case it that of a sextet made of two other sextets, 
which yields  $2 \times 6^3 = 432$ equations.
Like for the derivation of the multiplets in Eq.~\eqref{multiplets-final} themselves,
which proceeds along similar lines, we employ \texttt{Mathematica} for this task. 

In the following we  go through the columns in Table~\ref{tab:products} one by one.

\smallskip
\underline{$\sQ \otimes \sQ$}: Given two quartets 
\renewcommand{\arraystretch}{1.0}
\begin{equation}\label{S5-quartets}
   \mQ = \left[ \begin{array}{c} a \\ b \\ c \\ d \end{array}\right]\,, \quad
   \mQ' = \left[ \begin{array}{c} a' \\ b' \\ c' \\ d' \end{array}\right],
\end{equation}
their tensor product returns 16 elements which can again be arranged into multiplets.
Their dot product is a singlet,
\begin{equation}
   \mQ \cdot \mQ' = aa'+bb'+cc'+dd'\,,
\end{equation}
and the remaining 15 elements can be grouped into a quartet, a quintet and a sextet.
These are given explicitly in Table \ref{tab:4x4}.
To make the expressions more compact, we abbreviated 
\begin{equation} 
\begin{array}{rl}
   \Sigma_{ab} &= ab' + a'b\,, \\[1mm]
   \Delta_{ab} &= ab'-a'b \,,  
\end{array}\qquad
\begin{array}{rl}
   \Lambda_{ab} &=aa'+bb'\,, \\[1mm]
   \Psi_{ab} &=aa'-bb'\,.
\end{array}
\end{equation}
If the two input multiplets are identical, then $\Sigma_{ab} = 2ab$ and $\Delta_{ab}=0$, so one has for example $\mQ \ast \mQ = 0$.

\smallskip
\underline{$\sV \otimes \sV$}: From two quintets
\renewcommand{\arraystretch}{1.0}
\begin{equation}\label{S5-quintets}
   \mV = \left[ \begin{array}{c} e \\ f \\ g \\ h \\ j \end{array}\right]\,, \quad
   \mV' = \left[ \begin{array}{c} e' \\ f' \\ g' \\ h' \\j' \end{array}\right] 
\end{equation}
one obtains 25 elements.
The dot product is again singlet,
\begin{equation}
   \mV \cdot \mV' = ee'+ff'+gg'+hh'+jj'\,,
\end{equation}
and the remaining combinations form a quartet, antiquartet, quintet, antiquintet and sextet, which are collected in Table \ref{tab:4x4}.

\smallskip
\underline{$\sW \otimes \sW$}: From two sextets
\renewcommand{\arraystretch}{1.0}
\begin{equation}\label{S5-sextets}
    \mW = \left[ \begin{array}{c} u \\ v \\ w \\ x \\ y \\ z \end{array}\right]\,, \quad
    \mW' = \left[ \begin{array}{c} u' \\ v' \\ w' \\ x' \\y' \\ z' \end{array}\right] 
\end{equation}
one obtains a singlet and an antisinglet:
\begin{equation}
\begin{split}
   \mW \cdot \mW' &= uu'+vv'+ww'+xx'+yy'+zz'\,, \\
   \mW\times \mW' &= uz' + u'z - vy' -v'y + wx' + w'x\,.   
\end{split}
\end{equation}
Moreover, this gives a quartet, an antiquartet, two quintets, two antiquintets and a sextet, 
which are listed in Table \ref{tab:6x6}. Here we used the further abbreviations
\begin{equation} 
\begin{split}
   \Sigma_{ab,cd}^\pm &= \Sigma_{ab} \pm \Sigma_{cd}\,, \\ 
   \Delta_{ab,cd}^\pm &= \Delta_{ab} \pm \Delta_{cd}\,, \\
   \Lambda_{ab,cd}^\pm &= \Lambda_{ab} \pm \Lambda_{cd}\,, \\ 
   \Psi_{ab,cd}^\pm &= \Psi_{ab} \pm \Psi_{cd}\,.
\end{split}
\end{equation}
 
 \smallskip
\underline{$\sQ \otimes \sV$}: Combining a quartet $\mQ = \left[ a ,\, b,\, c,\, d \right]$ and a quintet $\mV = \left[ e ,\, f,\, g,\, h,\, j \right]$ gives 20 elements, 
which are grouped into a quartet, quintet, antiquintet and sextet in Table~\ref{tab:4x6}.
Here we abbreviated
\begin{equation} 
\begin{split}
   \Sigma_{ab}^{cd} &= a b + c d\,, \\
   \Delta_{ab}^{cd} &= a b - c d \,.
\end{split}
\end{equation}

\smallskip
\underline{$\sQ \otimes \sW$}: From a quartet $\mQ = \left[ a ,\, b,\, c,\, d \right]$ and a sextet $\mW = \left[ u ,\, v ,\, w ,\, x ,\, y ,\, z \right]$
one obtains 24 combinations, which form a quartet, antiquartet, quintet, antiquintet and a sextet (Table~\ref{tab:4x6}).

\smallskip
\underline{$\sV \otimes \sW$}: Finally, combining a quintet $\mV = \left[ e ,\, f,\, g,\, h,\, j \right]$ 
with a sextet $\mW = \left[ u ,\, v ,\, w ,\, x ,\, y ,\, z \right]$ gives 30 combinations,
which yield a quartet, antiquartet, quintet, antiquintet and two sextets (Table \ref{tab:5x6}).

\begin{table}[!p]
   \centering
   \begin{tabular}{l @{\;}  l } \hline\noalign{\smallskip}
      $\sQ\otimes\sQ$     &               \\ \noalign{\smallskip}\hline\noalign{\medskip}
	  $\mQ \vee \mQ'=$    &  $\left[\begin{array}{c}  
                            \Sigma_{ab} + \frac{1}{\sqrt{2}}\,\Sigma_{ac}+ \sqrt{\frac{3}{10}}\,\Sigma_{ad} \\[1mm]
                            \Psi_{ab} + \frac{1}{\sqrt{2}}\,\Sigma_{bc} + \sqrt{\frac{3}{10}}\,\Sigma_{bd}  \\[1mm] 
                            \frac{1}{\sqrt{2}}\,(\Lambda_{ab}- 2 cc') + \sqrt{\frac{3}{10}}\,\Sigma_{cd} \\[1mm] 
                            \sqrt{\frac{3}{10}}\,(\Lambda_{ab}+cc'-3 dd')
                            \end{array}\right]$  \\ \noalign{\medskip} 
     $\mQ \cup \mQ'=$    & $\left[\begin{array}{c}  
                           \Lambda_{ab}-2cc'-\sqrt{15}\,\Sigma_{cd}  \\[1mm] 
                            \sqrt{6}\,\Psi_{ab}-2\sqrt{3}\,\Sigma_{bc} \\[1mm] 
                            \sqrt{6}\,\Sigma_{ab}-2\sqrt{3}\,\Sigma_{ac}  \\[1mm] 
                            \sqrt{2}\,\Sigma_{ab}+\Sigma_{ac}-\sqrt{15}\,\Sigma_{ad} \\[1mm] 
                            \sqrt{2}\,\Psi_{ab}+\Sigma_{bc}-\sqrt{15}\,\Sigma_{bd} 
                            \end{array}\right]$ \\  \noalign{\medskip} 
     $\mQ \ast \mQ'=$    & $\big[\Delta_{ab},\, \Delta_{ac},\, \Delta_{ad},\, \Delta_{bc},\, \Delta_{bd},\, \Delta_{cd}\big]$  \\   
                           \noalign{\bigskip}\hline\noalign{\smallskip} 
        
      $\sV\otimes\sV$    &                                   \\ \noalign{\smallskip}\hline\noalign{\medskip}
     
      $\mV \vee \mV'=$   &  $\left[\begin{array}{c}  
                           \Sigma_{eg} + \frac{1}{\sqrt{3}}\,\Sigma_{eh} - \frac{1}{\sqrt{2}}\,(\Sigma_{fh} + \Sigma_{gj}) +\sqrt{\frac{2}{3}} \,\Sigma_{hj}  \\[1mm]
                           \Sigma_{ef} + \frac{1}{\sqrt{3}}\,\Sigma_{ej} - \frac{1}{\sqrt{2}}\,(\Sigma_{gh} - \Sigma_{fj}) +\sqrt{\frac{2}{3}} \,\Psi_{hj} \\[1mm]
                           \Sigma_{gh} + \Sigma_{fj}+ \frac{1}{\sqrt{3}}\,(\Lambda_{hj}-2ee') \\[1mm]
                           \frac{1}{\sqrt{5}}\,( 3\Lambda_{fj} - 2\Lambda_{eh} + 3 \Psi_{gj}-2jj' )   
                           \end{array}\right]$  \\ \noalign{\medskip} 
                           
      $\mV \wedge \mV'=$ &  $\left[\begin{array}{c} 
                            \Delta_{ef}-\sqrt{3}\,\Delta_{ej}+\frac{1}{\sqrt{2}}\,(\Delta_{fj} - \Delta_{gh}) \\[1mm]
                            -\Delta_{eg}+\sqrt{3}\,\Delta_{eh}+\frac{1}{\sqrt{2}}\,(\Delta_{fh}+ \Delta_{gj}) \\[1mm]
                            \Delta_{gj} - \Delta_{fh} +\sqrt{3}\,\Delta_{hj} \\[1mm]
                            \sqrt{5}\,\Delta_{fg}  
                            \end{array}\right]$  \\ \noalign{\medskip} 
                            
      $\mV \cup \mV'=$   &  $\left[\begin{array}{c}
                            \Lambda_{hj} - 2ee'-\sqrt{\frac{3}{4}}\,(\Sigma_{fj}+\Sigma_{gh}) \\[1mm]
                            \sqrt{\frac{3}{8}}\,(\Psi_{hj}-3\Psi_{fg})-\sqrt{\frac{3}{4}}\,\Sigma_{ej} \\[1mm]
                            -\sqrt{\frac{3}{4}}\,\Sigma_{eh}+\sqrt{\frac{3}{8}}\,(3\Sigma_{fg}+\Sigma_{hj}) \\[1mm]
                            -\sqrt{\frac{3}{4}}\,\Sigma_{eg}+\Sigma_{eh}+\sqrt{\frac{3}{8}}\,(\Sigma_{fh}+\Sigma_{gj})+\sqrt{2}\,\Sigma_{hj} \\[1mm]
                            -\sqrt{\frac{3}{4}}\,\Sigma_{ef}+\Sigma_{ej}+\sqrt{2}\Psi_{hj}-\sqrt{\frac{3}{8}}\,(\Sigma_{fj}-\Sigma_{gh})
                            \end{array}\right]$  \\ \noalign{\medskip} 
                            
      $\mV \cap \mV'=$   &  $\left[\begin{array}{c}
                            \Sigma_{fh}-\Sigma_{gj} \\[1mm]
                            \Sigma_{eh}+\frac{1}{\sqrt{2}}\,(\Sigma_{fg}-\Sigma_{hj}) \\[1mm]
                            \frac{1}{\sqrt{2}}\,(\Psi_{fg}+\Psi_{hj})-\Sigma_{ej} \\[1mm]
                            \Sigma_{ef}+\frac{1}{\sqrt{2}}\,(\Sigma_{gh}-\Sigma_{fj}) \\[1mm]
                            -\Sigma_{eg}-\frac{1}{\sqrt{2}}\,(\Sigma_{fh}+\Sigma_{gj}) 
                             \end{array}\right]$  \\ \noalign{\medskip} 
                             
      $\mV \ast \mV'=$   &  $\left[\begin{array}{c}
                            \Delta_{fh} -\Delta_{gj}+\frac{2}{\sqrt{3}}\,\Delta_{hj} \\[1mm]
                            -\Delta_{eg}-\frac{2}{\sqrt{3}}\,\Delta_{eh}+\frac{1}{\sqrt{2}}\,(\Delta_{fh}+\Delta_{gj}) \\[1mm]
                            -\sqrt{\frac{5}{6}}\left(\sqrt{2}\,\Delta_{eg}+\Delta_{fh}+\Delta_{gj}\right) \\[1mm]
                            -\Delta_{ef}-\frac{2}{\sqrt{3}}\,\Delta_{ej}+\frac{1}{\sqrt{2}}\,(\Delta_{gh}-\Delta_{fj}) \\[1mm]
                            -\sqrt{\frac{5}{6}}\,(\sqrt{2}\,\Delta_{ef}+\Delta_{gh}-\Delta_{fj}) \\[1mm]
                             \sqrt{\frac{5}{3}}\,(\Delta_{fj}+\Delta_{gh})
                             \end{array}\right]$  \\    
                             \noalign{\bigskip}\hline\noalign{\smallskip} 
                             
                     \end{tabular}
   \caption{ $\sQ\otimes\sQ$ and $\sV\otimes\sV$ multiplet products. }
   \label{tab:4x4}
\end{table}

\begin{table}[!p]
   \centering
   \begin{tabular}{c @{\;}  l } \hline\noalign{\smallskip}
      $\sW\otimes\sW$     &     \\ \noalign{\smallskip}\hline\noalign{\medskip}
      
      $\mW \vee \mW' =$      &  $\left[\begin{array}{c} 
                             -\sqrt{3}\,\Sigma^{+}_{uy,vz} - \sqrt{5}\,\Sigma^{-}_{ux,wz} + \sqrt{10}\,\Sigma^{+}_{vx,wy} \\[1mm]
                              \sqrt{3}\,\Sigma^-_{uw,xz} +\sqrt{5}\,\Sigma^{+}_{uv,yz} +\sqrt{10}\,\Lambda^{-}_{vw,xy} \\[1mm]
                              \sqrt{3}\,\Sigma^{+}_{xy,vw} + \sqrt{5}\,(2\Psi_{uz}-\Psi^{+}_{vw,xy}) \\[1mm]
                             2\sqrt{3}\,( \Lambda^{-}_{uv,yz}-\Psi_{wx} )
                             \end{array}\right]$  \\ \noalign{\medskip} 
                             
      $\mW \wedge \mW' =$    & $\left[\begin{array}{c} 
                             \sqrt{3}\,\Delta^{-}_{uv,yz}-\sqrt{5}\,\Delta^{+}_{uw,xz}-\sqrt{10}\,\Delta^{-}_{xy,vw} \\[1mm]
                             \sqrt{3}\,\Delta^{+}_{ux,wz}-\sqrt{5}\,\Delta^{-}_{uy,vz}-\sqrt{10}\,\Delta^{-}_{vy,wx} \\[1mm]
                             \sqrt{3}\,\Delta^{-}_{vx,wy}+\sqrt{5}\,(2\Delta_{uz}+\Delta^{+}_{vy,wx}) \\[1mm]
                             2\sqrt{3}\,(\Delta_{uz}-\Delta^{+}_{vy,wx})  
                             \end{array}\right]$  \\ \noalign{\medskip} 
                             
      $[\mW \cup \mW']_1 =$ & $\left[\begin{array}{c} 
                               2\Psi_{uz}-\Psi^+_{vw,xy}-\sqrt{15}\,\Sigma^+_{vw,xy} \\[1mm]
                               -\sqrt{6}\left( \sqrt{2}\,\Sigma^+_{uv,yz} - \Lambda^-_{vw,xy}  \right) \\[1mm]
                                \sqrt{6}\left( \sqrt{2}\,\Sigma^-_{ux,wz} + \Sigma^+_{vx,wy}  \right) \\[1mm]
                               -\Sigma^-_{ux,wz} + \sqrt{2}\,\Sigma^+_{vx,wy} + \sqrt{15}\,\Sigma^+_{uy,vz} \\[1mm]
                               \Sigma^+_{uv,yz}+\sqrt{2}\,\Lambda^-_{vw,xy}-\sqrt{15}\,\Sigma^-_{uw,xz}
                               \end{array}\right]$  \\ \noalign{\medskip} 
                               
      $[\mW \cup \mW']_2 =$  & $\left[\begin{array}{c} 
                               2\Lambda_{uz} - \Lambda^+_{vw,xy}  \\[1mm]
                               \begin{array}{rl}  \frac{1}{2\sqrt{2}} \,\big( &\!\!\! \sqrt{10}\,\Sigma^+_{uw,xz} + \sqrt{6}\,\Sigma^-_{uv,yz} \\
                                &\!\!\!  + \sqrt{5}\,\Sigma^-_{vw,xy} - \sqrt{3}\,\Psi^-_{vw,xy} \big) \end{array} \\[4mm]
                               \begin{array}{rl} -\frac{1}{2\sqrt{2}} \,\big( &\!\!\! \sqrt{10}\,\Sigma^-_{uy,vz} + \sqrt{6}\,\Sigma^+_{ux,wz} \\
                               &\!\!\! - \sqrt{5}\,\Sigma^+_{vy,wx} + \sqrt{3}\,\Sigma^-_{vx,wy}  \big) \end{array}\\[4mm]
                               -\Sigma^+_{ux,wz} + \sqrt{2}\,\Sigma^-_{vx,wy} \\[1mm]
                                \Sigma^-_{uv,yz} + \sqrt{2}\,\Psi^-_{vw,xy}
                               \end{array}\right]$  \\ \noalign{\medskip} 

      $[\mW \cap \mW']_1 =$  & $\left[\begin{array}{c} 
                               -(2\Delta_{uz}+\Delta^+_{vy,wx})+\sqrt{15}\,\Delta^-_{vx,wy} \\[1mm]
                               -\sqrt{6}\left( \sqrt{2}\,\Delta^-_{uy,vz} - \Delta^-_{vy,wx}  \right) \\[1mm]
                               -\sqrt{6}\left( \sqrt{2}\,\Delta^+_{uw,xz} + \Delta^-_{vw,xy}  \right) \\[1mm]
                               \Delta^+_{uw,xz} - \sqrt{2}\,\Delta^-_{vw,xy} + \sqrt{15}\,\Delta^-_{uv,yz} \\[1mm]
                               \Delta^-_{uy,vz}+\sqrt{2}\,\Delta^-_{vy,wz}+\sqrt{15}\,\Delta^+_{ux,wz}
                               \end{array}\right]$  \\ \noalign{\medskip} 

      $[\mW \cap \mW']_2 =$  & $\left[\begin{array}{c} 
                               2\Sigma_{uz} + \Sigma^-_{vy,wx}  \\[1mm]
                               \begin{array}{rl} \frac{1}{2\sqrt{2}}\,\big( &\!\!\! \sqrt{10}\,\Sigma^+_{ux,wz}-\sqrt{6}\,\Sigma^-_{uy,vz} \\
                                &\!\!\! + \sqrt{5}\,\Sigma^-_{vx,wy} + \sqrt{3}\,\Sigma^+_{vy,wx} \big) \end{array} \\[4mm]
                               \begin{array}{rl} \frac{1}{2\sqrt{2}}\,\big( &\!\!\! \sqrt{10}\,\Sigma^-_{uv,yz}-\sqrt{6}\,\Sigma^+_{uw,xz} \\
                               &\!\!\! - \sqrt{5}\,\Psi^-_{vw,xy} - \sqrt{3}\,\Sigma^-_{vw,xy} \big) \end{array}\\[4mm]
                               -\Sigma^+_{uw,xz} + \sqrt{2}\,\Sigma^-_{vw,xy} \\[1mm]
                               -\Sigma^-_{uy,vz} - \sqrt{2}\,\Sigma^+_{vy,wx}
                               \end{array}\right]$  \\ \noalign{\medskip} 

      $\mW \ast \mW' =$      &  $\left[\begin{array}{c} 
                                \Delta^{+}_{vx,wy} \\[1mm]
                                -\Delta^{-}_{ux,wz} \\[1mm] 
                                -\Delta^{+}_{uy,vz} \\[1mm]
                                 \Delta^{+}_{uv,yz} \\[1mm]
                                  \Delta^{-}_{uw,xz} \\[1mm]
                                   \Delta^{+}_{vw,xy} \end{array}\right]$  \\    
                             \noalign{\bigskip}\hline\noalign{\smallskip} 
   \end{tabular}
   \caption{ $\sW\otimes\sW$ multiplet products. Here we abbreviated $\Sigma^{\pm}_{ab,cd} = \Sigma_{ab} \pm \Sigma_{cd}$,
             and likewise for $\Delta^{\pm}_{ab,cd}$, $\Lambda^{\pm}_{ab,cd}$ and $\Psi^\pm_{ab,cd}$. }
   \label{tab:6x6}
\end{table}

   \begin{table}[p]
      \centering
      \begin{tabular}{l @{\,}  l } \hline\noalign{\smallskip}
         $\sQ\otimes\sV$      &                                      \\ \noalign{\smallskip}\hline\noalign{\medskip}
         
         $\mQ\vee\mV = $        & $\left[\begin{array}{c} 
                                 \Sigma_{ae}^{ch}+\sqrt{2}\,\Sigma_{aj}^{bh}+\sqrt{6}\,\Sigma_{af}^{bg}-2\sqrt{3}\,cg-\sqrt{15}\,dh, \\[1mm]
                                 \Sigma_{be}^{cj}+\sqrt{2}\,\Delta_{ah}^{bj}+\sqrt{6}\,\Delta_{ag}^{bf}-2\sqrt{3}\,cf-\sqrt{15}\,dj \\[1mm]
                                 \Sigma_{ah}^{bj}-2ce -2\sqrt{3}\,\Sigma_{ag}^{bf}-\sqrt{15}\,de \\[1mm] 
                                 -\sqrt{15}\,(\Sigma_{ah}^{bj}+ce)  
                                 \end{array}\right]$  \\ \noalign{\medskip} 

         $\mQ\cup\mV= $         & $\left[\begin{array}{c} 
                                  \sqrt{10}\,(\Sigma_{ah}^{bj}-2ce) + \sqrt{30}\,\Sigma_{ag}^{bf}-2\sqrt{6}\,de \\[1mm] 
                                  -\sqrt{15}\,\Delta_{ah}^{bj} + \sqrt{30}\,\Sigma_{be}^{cj}+3\sqrt{6}\,df \\[1mm] 
                                  -\sqrt{15}\,\Sigma_{aj}^{bh} + \sqrt{30}\,\Sigma_{ae}^{ch} + 3\sqrt{6}\,dg \\[1mm]
                                  
                                  \begin{array}{rl} \big( &\!\!\! \sqrt{10}\,\Sigma_{ae}^{ch} - \sqrt{15}\,\Sigma_{af}^{bg} + \sqrt{30}\,cg \\
                                  & \quad + 2\sqrt{5}\,\Sigma_{aj}^{bh}- 2\sqrt{6}\,dh \big) \end{array}   \\[3mm] 
                                  
                                  \begin{array}{rl} \big( &\!\!\! \sqrt{10}\,\Sigma_{be}^{cj} - \sqrt{15}\,\Delta_{ag}^{bf} + \sqrt{30}\,cf \\
                                   & \quad + 2\sqrt{5}\,\Delta_{ah}^{bj} - 2\sqrt{6}\,dj \big) \end{array}    
                                   
                                   \end{array}\right]$  \\ \noalign{\medskip} 

         $\mQ\cap\mV = $         & $\left[\begin{array}{c} 
                                   \sqrt{2}\,\Delta_{af}^{bg} - \sqrt{6}\,\Delta_{aj}^{bh} \\[1mm] 
                                    \Sigma_{aj}^{bh} - \sqrt{2}\,(\Delta_{ae}^{ch}+2ch) + \sqrt{10}\,dg \\[1mm] 
                                    -\Delta_{ah}^{bj}+ \sqrt{2}\,\Sigma_{be}^{cj}-\sqrt{10}\,df\\[1mm] 
                                    \Delta_{ag}^{bf}- \sqrt{6}\,\Delta_{be}^{cj} + \sqrt{2}\,cf\\[1mm]  
                                    -\Sigma_{af}^{bg} + \sqrt{6}\,\Delta_{ae}^{ch}- \sqrt{2}\,cg
                                    \end{array}\right]$  \\ \noalign{\medskip} 

         $\mQ\ast\mV = $         & $\left[\begin{array}{c} 
                                   -\sqrt{10}\,\Delta_{bh}^{aj} + \sqrt{30}\,\Delta_{af}^{bg} \\[1mm] 
                                   \sqrt{10}\,\Delta_{ae}^{ch} +\sqrt{15}\,\Sigma_{af}^{bg} + \sqrt{30}\,c g \\[1mm]  
                                   \Sigma_{af}^{bg} + \sqrt{6}\,\Sigma_{ae}^{ch} + 2\sqrt{3}\,\Sigma_{aj}^{bh}- \sqrt{2}\,cg + 2\sqrt{10}\, dh \\[1mm] 
                                   \sqrt{10}\,\Delta_{be}^{cj}+ \sqrt{15}\,\Delta_{ag}^{bf}+ \sqrt{30}\,cf\\[1mm] 
                                   \Delta_{ag}^{bf} + 2\sqrt{3}\,\Delta_{ah}^{bj} + \sqrt{6}\,\Sigma_{be}^{cj}- \sqrt{2}\,cf+ 2\sqrt{10}\,dj \\[1mm] 
                                   -\sqrt{2}\,\Sigma_{ag}^{bf} + \sqrt{6}\,\Sigma_{ah}^{bj}- 2\sqrt{6}\,ce + 2\sqrt{10}\,de
                                    \end{array}\right]$  \\    
                             \noalign{\medskip}\hline\noalign{\smallskip} 

         $\sQ\otimes\sW$      &                                   \\ \noalign{\smallskip}\hline\noalign{\medskip}
         
         $\mQ\vee\mW =$         &  $\left[\begin{array}{c} 
                                   bu+cv+dw \\
                                   -au+cx+dy \\
                                   -av-bx+dz \\
                                   -aw-by-cz
                                   \end{array}\right]$,   
                                   
         $\mQ\wedge\mW = \left[\begin{array}{c} 
                                    bz-cy+dx \\
                                    -az+cw-dv \\
                                     ay-bw+du \\
                                     -ax+bv-cu
                                     \end{array}\right]$  \\ \noalign{\medskip} 

         $\mQ\cup\mW=$         &  $\left[\begin{array}{c} 
                                  \sqrt{6}\,(\Sigma_{aw}^{by}-2cz) + \sqrt{10}\,(\Sigma_{av}^{bx}+2dz) \\[1mm]  
                                  \Delta_{aw}^{by} - \sqrt{2}\,\Sigma_{bz}^{cy} + \sqrt{15}\,\Delta_{av}^{bx} + \sqrt{30}\,\Sigma_{au}^{cx} \\[1mm]  
                                  \Sigma_{ay}^{bw} - \sqrt{2}\,\Sigma_{az}^{cw} + \sqrt{15}\,\Sigma_{ax}^{bv} - \sqrt{30}\,\Delta_{bu}^{cv} \\[1mm]  
                                  \sqrt{6}\,\Sigma_{az}^{cw} - \sqrt{10}\,(\Sigma_{bu}^{cv} -2dw) + 2\sqrt{3}\,\Sigma_{ay}^{bw} \\[1mm]
                                  \sqrt{6}\,\Sigma_{bz}^{cy} + \sqrt{10}\,(\Delta_{au}^{cx} +2dy) + 2\sqrt{3}\,\Delta_{aw}^{by}
                                  \end{array}\right]$  \\ \noalign{\medskip} 

         $\mQ\cap\mW=$         & $\left[\begin{array}{c} 
                                  \sqrt{6}\,(\Delta_{ax}^{bv}-2cu) - \sqrt{10}\,(\Delta_{ay}^{bw}-2du) \\[1mm]  
                                  \Sigma_{ax}^{bv} - \sqrt{2}\,\Delta_{bu}^{cv} - \sqrt{15}\,\Sigma_{ay}^{bw} + \sqrt{30}\,\Sigma_{az}^{cw} \\[1mm]  
                                  -\Delta_{av}^{bx} - \sqrt{2}\,\Sigma_{au}^{cx} + \sqrt{15}\,\Delta_{aw}^{by}- \sqrt{30}\,\Sigma_{bz}^{cy} \\[1mm] 
                                  \sqrt{6}\,\Sigma_{au}^{cx} - \sqrt{10}\,(\Delta_{bz}^{cy} - 2dx) - 2\sqrt{3}\,\Delta_{av}^{bx}  \\[1mm] 
                                  \sqrt{6}\,\Delta_{bu}^{cv} + \sqrt{10}\,(\Delta_{az}^{cw} - 2dv) + 2\sqrt{3}\,\Sigma_{ax}^{bv}
                                  \end{array}\right]$  \\ \noalign{\medskip} 

         $\mQ\ast\mW=$         &  $\left[\begin{array}{c} 
                                  \sqrt{3}\,(\Delta_{ay}^{bw} -2du) + \sqrt{5}\,(\Delta_{ax}^{bv} - 2cu) \\[1mm] 
                                  \sqrt{3}\,(\Delta_{az}^{cw} -2dv) - \sqrt{5}\,\Delta_{bu}^{cv} - \sqrt{10}\,\Sigma_{ax}^{bv} \\[1mm] 
                                  -\sqrt{3}\,(\Sigma_{bu}^{cv}-2dw) -\sqrt{5}\,\Sigma_{az}^{cw} - \sqrt{10}\,\Sigma_{ay}^{bw} \\[1mm] 
                                  \sqrt{3}\,(\Delta_{bz}^{cy} -2dx)+ \sqrt{5}\,\Sigma_{au}^{cx} - \sqrt{10}\,\Delta_{av}^{bx} \\[1mm] 
                                  \sqrt{3}\,(\Delta_{au}^{cx}+2dy) - \sqrt{5}\,\Sigma_{bz}^{cy} - \sqrt{10}\,\Delta_{aw}^{by} \\[1mm] 
                                  \sqrt{3}\,(\Sigma_{av}^{bx}+2dz)- \sqrt{5}\,(\Sigma_{aw}^{by} - 2cz)
                                  \end{array}\right]$  \\    
                             \noalign{\medskip}\hline\noalign{\smallskip} 

      \end{tabular}
      \caption{$\sQ\otimes\sV$ and $\sQ\otimes\sW$ multiplet products. }
      \label{tab:4x6}
   \end{table}

\section{Relations between Lorentz invariants}\label{sec:li-relations}   

In this appendix we collect the rotation matrices between the different sets of Lorentz invariants defined in Sec.~\ref{5pf}
for convenience.
We combine the Jacobi variables from Eq.~\eqref{li-variables-jacobi}, the Mandelstam variables from Eq.~\eqref{mandelstams},
and the variables defined in Eq.~\eqref{uv-var} into the following vectors:
\begin{equation*}
\begin{split}
    \vect{v}_1 &= \{ p^2, \, q^2, \, k^2, \, l^2, \, \omega_1, \, \omega_2, \, \omega_3, \, \omega_4, \, \omega_5, \, \omega_6 \}\,, \\
    \vect{v}_2 &= \{ s_{12}, \,s_{13}, \,s_{14}, \,s_{15}, \,s_{23}, \,s_{24}, \,s_{25},\, s_{34}, \,s_{35},\, s_{45} \}\,, \\
    \vect{v}_3 &= \{ \mQ\cdot \mQ, \,u_1, \,u_2, \,u_3, \,u_4, \,v_1, \,v_2,\, v_3, \,v_4,\, v_5 \}\,.
\end{split}
\end{equation*}
Writing $\vect{v}_1 = \mathsf{M} \,\vect{v}_2$ and $\vect{v}_2 = \mathsf{M}' \,\vect{v}_3$,
the  rotation matrices are given by
\begin{equation*} \renewcommand{\arraystretch}{1.0}
\begin{split}
    \mathsf{M} &= \frac{1}{6}\left( \begin{array}{rrrrrrrrrr} 
                                                    10 & 4  &  4 &  4 &  4 &  4 &  4 & -8 & -8 & -8  \\ 
                                                     2 & 8  & -4 & -4 &  8 & -4 & -4 &  8 &  8 & -8 \\
                                                    -3 & -3 & 9  & -6 & -3 &  9 & -6 &  9 & -6 & 10 \\
                                                    -5 & -5 & -5 & 10 & -5 & -5 & 10 & -5 & 10 & 10 \\
                                                     0 &  4 &  8 &  8 & -4 & -8 & -8 &  0 &  0 &  0 \\
                                                     0 &  5 &  1 &  4 & -5 & -1 & -4 &  0 &  0 &  0 \\
                                                    10 & -5 &  1 &  4 & -5 &  1 &  4 & -2 & -8 &  0 \\
                                                     0 &  3 &  3 &  0 & -3 & -3 &  0 &  0 &  0 &  0 \\
                                                     6 & -3 &  3 &  0 & -3 &  3 &  0 & -6 &  0 &  0 \\
                                                     6 &  6 & -6 &  0 &  6 & -6 &  0 & -6 &  0 &  0    \end{array}\right), \\
    \mathsf{M}' &= \frac{1}{10}\left( \begin{array}{rrrrrrrrrr} 
                                                     1 &  3 & -2 & -2 & -2 & 15 &  0 &  0 &  0 & -5  \\ 
                                                     1 & -2 &  3 &  3 &  3 & -5 &  5 & -5 &  0 &  5 \\
                                                     1 & -2 &  3 &  3 &  3 & -5 & -5 & -5 &  0 &  5 \\
                                                     1 &  3 & -2 & -2 & -2 & -5 &  0 & 10 &  0 & -5 \\
                                                     1 & -2 & -7 &  3 &  3 & -5 & -5 &  5 & -5 &  0 \\
                                                     1 & -2 &  3 & -7 &  3 & -5 &  5 &  5 &  5 &  0 \\
                                                     1 & -7 &  8 &  8 & -2 & -5 &  0 &-10 &  0 &  5 \\
                                                     1 &  3 & -2 & -2 & -2 &  5 &  0 &  0 &  0 & -5 \\
                                                     1 &  3 & -2 & -2 & -2 &  5 &  0 &  0 &  5 &  0 \\
                                                     1 &  3 & -2 & -2 & -2 &  5 &  0 &  0 & -5 &  0    \end{array}\right).
\end{split}
\end{equation*} 

We also give the relations for the variables $x_i = p_i^2$ and $x_{ij} = p_i \cdot p_j$, where $p_1 \dots p_5$ are the external momenta.
Writing $\vect{v}_4 = \{ x_1, \,x_2, \,x_3, \,x_4, \,x_5, \, x_{12}, \, x_{13}, \, x_{14}, \, x_{15}, \, x_{23},$ $  x_{24},  x_{25}, \,x_{34},\, x_{35}, \, x_{45} \}$
and $\vect{v}_4 = \mathsf{M}'' \,\vect{v}_2$, the rotation matrix is
\begin{equation*} \renewcommand{\arraystretch}{1.0}
\begin{split}
    \mathsf{M}'' &= -\frac{1}{6}\left( \begin{array}{rrrrrrrrrr} 
                                                    -4 & -4 & -4 & -4 &  2 &  2 &  2 &  2 &  2 &  2  \\ 
                                                    -4 &  2 &  2 &  2 & -4 & -4 & -4 &  2 &  2 &  2 \\
                                                     2 & -4 &  2 &  2 & -4 &  2 &  2 & -4 & -4 &  2 \\
                                                     2 &  2 & -4 &  2 &  2 & -4 &  2 & -4 &  2 & -4 \\
                                                     2 &  2 &  2 & -4 &  2 &  2 & -4 &  2 & -4 & -4 \\
                                                     1 &  1 &  1 &  1 &  1 &  1 &  1 & -2 & -2 & -2 \\
                                                     1 &  1 &  1 &  1 &  1 & -2 & -2 &  1 &  1 & -2 \\
                                                     1 &  1 &  1 &  1 & -2 &  1 & -2 &  1 & -2 &  1 \\
                                                     1 &  1 &  1 &  1 & -2 & -2 &  1 & -2 &  1 &  1 \\
                                                     1 &  1 & -2 & -2 &  1 &  1 &  1 &  1 &  1 & -2  \\
                                                     1 & -2 &  1 & -2 &  1 &  1 &  1 &  1 & -2 &  1  \\
                                                     1 & -2 & -2 &  1 &  1 &  1 &  1 & -2 &  1 &  1  \\
                                                    -2 &  1 &  1 & -2 &  1 &  1 & -2 &  1 &  1 &  1  \\
                                                    -2 &  1 & -2 &  1 &  1 & -2 &  1 &  1 &  1 &  1  \\
                                                    -2 & -2 &  1 &  1 & -2 &  1 &  1 &  1 &  1 &  1    \end{array}\right).
\end{split}
\end{equation*}    
    
   \begin{table*}[t]
      \centering
      \begin{tabular}{l@{\;}  l } \hline\noalign{\smallskip}
         $\sV\otimes\sW$      &                                   \\ \noalign{\smallskip}\hline\noalign{\medskip}
        
         $\mV\vee\mW=$         & $\left[\begin{array}{c} 
                                 \Sigma_{fw}^{gy} + 2\sqrt{3}\,\Sigma_{hy}^{jw} + \sqrt{6}\,\Sigma_{ew}^{hz} + \sqrt{10}\,\Sigma_{ev}^{ju}
                                 + \sqrt{15}\,\Sigma_{fv}^{gx}+ \sqrt{30}\,fu - \sqrt{2}\,gz \\[1mm]  
                                 -\Delta_{fy}^{gw} + 2\sqrt{3}\,\Delta_{hw}^{jy} + \sqrt{6}\,\Sigma_{ey}^{jz} + \sqrt{10}\,\Delta_{ex}^{hu}
                                 -\sqrt{15}\,\Delta_{fx}^{gv} - \sqrt{30}\,gu - \sqrt{2}\,fz \\[1mm]  
                                 -\sqrt{2}\,\Sigma_{fy}^{gw} + \sqrt{6}\,(\Sigma_{hw}^{jy}-2ez) - \sqrt{10}\,\Sigma_{hv}^{jx} + \sqrt{30}\,\Sigma_{fx}^{gv} \\[1mm]  
                                 2\sqrt{10}\,(ez+hw+jy)
                                  \end{array}\right]$  \\ \noalign{\medskip} 
                                  
         $\mV\wedge\mW =$      & $\left[\begin{array}{c} 
                               \Delta_{fx}^{gv} - 2\sqrt{3}\,\Delta_{hv}^{jx}+ \sqrt{6}\,\Sigma_{ex}^{hu} 
                               - \sqrt{10}\,\Delta_{ey}^{jz} - \sqrt{15}\,\Delta_{fy}^{gw} + \sqrt{30}\,fz - \sqrt{2}\,gu \\[1mm]  
                               \Sigma_{fv}^{gx} + 2\sqrt{3}\,\Sigma_{hx}^{jv}- \sqrt{6}\,\Delta_{ev}^{ju} 
                               + \sqrt{10}\,\Delta_{ew}^{hz} - \sqrt{15}\,\Sigma_{fw}^{gy} - \sqrt{30}\,gz - \sqrt{2}\,fu \\[1mm]   
                              \sqrt{2}\,\Delta_{fv}^{gx} + \sqrt{6}\,\,(\Delta_{hx}^{jv} - 2eu) + \sqrt{10}\,\Delta_{hy}^{jw} + \sqrt{30}\,\Delta_{fw}^{gy} \\[1mm]  
                               2\sqrt{10}\,(eu+hx-jv)
                               \end{array}\right]$  \\ \noalign{\medskip} 
                               
         $\mV\cup\mW=$         & $\left[\begin{array}{c} 
                                 2\sqrt{2}\,\Sigma_{hv}^{jx} + \sqrt{6}\,\Sigma_{fx}^{gv} + \sqrt{10}\,\Sigma_{fy}^{gw}\\[1mm]  
                                 -\sqrt{3}\,\Delta_{hv}^{jx} - \sqrt{5}\,\Delta_{jy}^{hw}- \sqrt{6}\,\Sigma_{ex}^{hu}- \sqrt{10}\,\Sigma_{ey}^{jz} \\[1mm]  
                                 -\sqrt{3}\,\Sigma_{hx}^{jv} + \sqrt{5}\,\Sigma_{hy}^{jw}- \sqrt{6}\,\Delta_{ev}^{ju} - \sqrt{10}\,\Sigma_{ew}^{hz} \\[1mm]  
                                  -2\sqrt{2}\,\Sigma_{ev}^{ju} + \sqrt{3}\,\Sigma_{fv}^{gx}  - \sqrt{5}\,\Sigma_{fw}^{gy}  + \sqrt{6}\,fu + \sqrt{10}\,gz\\[1mm]   
                                  -2\sqrt{2}\,\Delta_{ex}^{hu} - \sqrt{3}\,\Delta_{fx}^{gv}  + \sqrt{5}\,\Delta_{fy}^{gw} - \sqrt{6}\,gu + \sqrt{10}\,fz
                               \end{array}\right]$  \\ \noalign{\medskip} 
                               
         $\mV\cap\mW=$         & $\left[\begin{array}{c} 
                                 2\sqrt{2}\,\Delta_{hy}^{jw} - \sqrt{6}\,\Delta_{fw}^{gy} + \sqrt{10}\,\Delta_{fv}^{gx}\\[1mm]  
                                 -\sqrt{3}\,\Sigma_{hy}^{jw} - \sqrt{5}\,\Sigma_{hx}^{jv} + \sqrt{6}\,\Sigma_{ew}^{hz}- \sqrt{10}\,\Delta_{ev}^{ju} \\[1mm]  
                                 \sqrt{3}\,\Delta_{hw}^{jy} + \sqrt{5}\,\Delta_{hv}^{jx}- \sqrt{6}\,\Sigma_{ey}^{jz} + \sqrt{10}\,\Sigma_{ex}^{hu} \\[1mm]  
                                  -2\sqrt{2}\,\Delta_{ey}^{jz} + \sqrt{3}\,\Delta_{fy}^{gw}  + \sqrt{5}\,\Delta_{fx}^{gv}  - \sqrt{6}\,fz - \sqrt{10}\,gu\\[1mm]   
                                   2\sqrt{2}\,\Delta_{ew}^{hz} + \sqrt{3}\,\Sigma_{fw}^{gy}  + \sqrt{5}\,\Sigma_{fv}^{gx} + \sqrt{6}\,gz - \sqrt{10}\,fu
                               \end{array}\right]$  \\ \noalign{\medskip} 
                               
         $[\mV\ast\mW]_{1}=$   & $\left[\begin{array}{c} 
                                  \Delta_{hx}^{jv} -2 e u + 2\sqrt{3}\,\Delta_{fv}^{gx} - \sqrt{15}\,\Delta_{hy}^{jw}\\[1mm]  
                                 \Delta_{ev}^{ju} - \sqrt{2}\,\Sigma_{hx}^{jv} - \sqrt{6}\,\Sigma_{fv}^{gx} + \sqrt{15}\,\Delta_{ew}^{hz} + 2\sqrt{3}\,fu \\[1mm]  
                                 -\Sigma_{ew}^{hz}  - \sqrt{2}\,\Sigma_{hy}^{jw} - \sqrt{6}\,\Sigma_{fw}^{gy} + \sqrt{15}\,\Sigma_{ev}^{ju} + 2\sqrt{3}\,gz \\[1mm]  
                                 \Sigma_{ex}^{hu} - \sqrt{2}\,\Delta_{hv}^{jx}+ \sqrt{6}\,\Delta_{fx}^{gv}  + \sqrt{15}\,\Delta_{ey}^{jz}- 2\sqrt{3}\,g u \\[1mm]   
                                  -\Sigma_{ey}^{jz} - \sqrt{2}\,\Delta_{hw}^{jy} + \sqrt{6}\,\Delta_{fy}^{gw}+ \sqrt{15}\,\Delta_{ex}^{hu}  + 2\sqrt{3}\,fz \\[1mm]  
                                  -\Sigma_{hw}^{jy} +2ez + 2\sqrt{3}\,\Sigma_{fy}^{gw} - \sqrt{15}\,\Sigma_{hv}^{jx}
                               \end{array}\right]$  \\ \noalign{\medskip} 
                               
         $[\mV\ast\mW]_{2}=$  & $\left[\begin{array}{c} 
                                  2\sqrt{2}\,(\Delta_{hx}^{jv} -2eu) -\sqrt{6}\,\Delta_{fv}^{gx}-\sqrt{10}\,\Delta_{fx}^{gy}\\[1mm]  
                                  2\sqrt{2}\,\Delta_{ev}^{ju} -4 \Sigma_{hx}^{jv} + \sqrt{3}\,\Sigma_{fv}^{gx} - \sqrt{5}\,\Sigma_{fw}^{gy} - \sqrt{6}\,fu -\sqrt{10}\,gz \\[1mm]  
                                  2\sqrt{2}\,\Sigma_{ew}^{hz} +4 \Sigma_{hy}^{jw} - \sqrt{3}\,\Sigma_{fw}^{gy} - \sqrt{5}\,\Sigma_{fv}^{gx} + \sqrt{6}\,gz -\sqrt{10}\,fu \\[1mm]  
                                  2\sqrt{2}\,\Sigma_{ex}^{hu} -4 \Delta_{hv}^{jx} - \sqrt{3}\,\Delta_{fx}^{gv} + \sqrt{5}\,\Delta_{fy}^{gw} + \sqrt{6}\,gu -\sqrt{10}\,fz \\[1mm]   
                                  2\sqrt{2}\,\Sigma_{ey}^{jz} +4 \Delta_{hw}^{jy} + \sqrt{3}\,\Delta_{fy}^{gw} + \sqrt{5}\,\Delta_{fx}^{gv} + \sqrt{6}\,fz +\sqrt{10}\,gu \\[1mm]  
                                  2\sqrt{2}\,(\Sigma_{hw}^{jy} -2ez) + \sqrt{6}\,\Sigma_{fy}^{gw} - \sqrt{10}\,\Sigma_{fx}^{gv}
                               \end{array}\right]$  \\    
                             \noalign{\bigskip}\hline\noalign{\smallskip} 
      \end{tabular}
      \caption{ $\sV\otimes\sW$ multiplet products.}
      \label{tab:5x6}
   \end{table*}

\clearpage

\section{Dimensional constraint}\label{sec:constraint}
   
Here we return to the dimensional constraint discussed in Sec.~\ref{5bwf}.
Suppose we have a total momentum $P$ and four relative momenta, which we denote by $q_i$ with $i=1\dots 4$.
This gives 15 Lorentz invariants, namely four $q_i^2$, six $\omega_{ij} =  q_i \cdot q_j$, four $\eta_i = q_i\cdot P$, and $P^2$.
To derive the constraint, we cast the four-vectors in a given reference frame, 
\begin{equation*} \renewcommand{\arraystretch}{1.0}
   \begin{split}
      P &= iM\left[ \begin{array}{c} 0 \\ 0 \\ 0 \\ 1 \end{array}\right]\!, \quad
      q_1 = \sqrt{q_1^2}\left[ \begin{array}{c} 0 \\ 0 \\ \bar{z}_1 \\ z_1 \end{array}\right]\!, \quad
      q_2 = \sqrt{q_2^2}\left[ \begin{array}{c} 0 \\ \bar{z}_2\,\bar{y}_1 \\ \bar{z}_2\,y_1 \\ z_2 \end{array}\right]\!, \\
      q_3 &= \sqrt{q_3^2}\left[ \begin{array}{c} \bar{z}_3\,\bar{y}_2\,\sin\phi \\ \bar{z}_3\,\bar{y}_2\,\cos\phi \\ \bar{z}_3\,y_2 \\ z_3 \end{array}\right]\!, \quad
      q_4 = \sqrt{q_4^2}\left[ \begin{array}{c} \bar{z}_4\,\bar{y}_3\,\sin\psi \\ \bar{z}_4\,\bar{y}_3\,\cos\psi \\ \bar{z}_4\,y_3 \\ z_4 \end{array}\right]\!.
   \end{split}
\end{equation*}
Here we abbreviated $\bar{a} = \sqrt{1-a^2}$, such that $\bar{a}^2 + a^2 = 1$.
In this case, the 14 independent variables are $P^2$, the four $q_i^2$, the four $z_i$, the three $y_i$, and the two angles $\phi$ and $\psi$.
We can then extract the four  $z_i$ from  the Lorentz-invariant relations $z_i = \hat{q_i}\cdot \hat{P}$, 
where $\hat{p}$ denotes the unit vector corresponding to a four-momentum $p$. 
For the remaining variables, 
we apply the Gram-Schmidt procedure and orthonormalize the $q_i$ with respect to the total momentum $P$. This yields the unit vectors
\begin{equation*} \renewcommand{\arraystretch}{1.0}
   \begin{split}
      \left[ \begin{array}{c} 0 \\ 0 \\ 1 \\ 0 \end{array}\right]\!, \quad
      \left[ \begin{array}{c} 0 \\ \bar{y}_1 \\ y_1 \\ 0 \end{array}\right]\!, \quad
      \left[ \begin{array}{c} \bar{y}_2\,\sin\phi \\ \bar{y}_2\,\cos\phi \\ y_2 \\ 0 \end{array}\right]\!, \quad
      \left[ \begin{array}{c} \bar{y}_3\,\sin\psi \\ \bar{y}_3\,\cos\psi \\ y_3 \\ 0 \end{array}\right]\!,
   \end{split}
\end{equation*}
which we call $n_0$, $n_1$, $n_2$, $n_3$, respectively.
The three variables $y_i$  then follow from the Lorentz-invariant relations $y_i = n_0\cdot n_i$.
If we further orthonormalize the momenta $n_1$, $n_2$, $n_3$ with respect to $n_0$, we get
\begin{equation*} \renewcommand{\arraystretch}{1.0}
   \begin{split}
      n_1' = \left[ \begin{array}{c} 0 \\ 1 \\ 0 \\ 0 \end{array}\right]\!, \quad
      n_2' = \left[ \begin{array}{c} \sin\phi \\ \cos\phi \\ 0 \\ 0 \end{array}\right]\!, \quad
      n_3' = \left[ \begin{array}{c} \sin\psi \\ \cos\psi \\ 0 \\ 0 \end{array}\right]\!,
   \end{split}
\end{equation*}
which gives the two remaining variables
\begin{equation}
   Y_1 = n_1'\cdot n_2'= \cos\phi\,, \quad
   Y_2 = n_1'\cdot n_3' = \cos\psi\,.
\end{equation}
The leftover Lorentz invariant
\begin{equation*}
   \begin{split}
      Y_3 = n_2'\cdot n_3' &= \cos\phi\,\cos\psi + \sin\phi\,\sin\psi = \cos(\phi-\psi) 
   \end{split}
\end{equation*}
is no longer independent because it satisfies
\begin{equation}\label{dim-constraint-1}
   \begin{split}
      Y_3 &= Y_1\,Y_2 + \sqrt{1-Y_1^2}\sqrt{1-Y_2^2} \\
      &\Rightarrow \quad Y_1^2 + Y_2^2 + Y_3^2 = 1 + 2Y_1\,Y_2\,Y_3\,,
   \end{split}
\end{equation}
which is the desired constraint equation.

Expressed in terms of   the original variables $q_i^2$, $\omega_{ij}$, $\eta_i$ and $P^2$, Eq.~\eqref{dim-constraint-1} turns into  the permutation-group invariant constraint 
\begin{equation}\label{dim-constraint}
   \begin{split}
       & \left[\left( \omega_{12}\,\omega_{34} + \omega_{13}\,\omega_{24}+ \omega_{14}\,\omega_{23}\right)^2  - q_1^2\,q_2^2\,q_3^2\,q_4^2\right]P^2\\[1mm]
       &+\sum_i \eta_i^2 \left[ q_j^2\,q_k^2\,q_l^2 - q_j^2\,\omega_{kl}^2 - q_k^2\,\omega_{jl}^2 - q_l^2\,\omega_{jk}^2 \right] \\
       &+ 2 \sum_i \left( \eta_i^2 - q_i^2\,P^2\right) \omega_{jk}\,\omega_{jl}\,\omega_{kl} \\
       &+ \sum_{i<j}  \eta_i\,\eta_j\,\Big[ 2q_k^2\left( \omega_{il}\,\omega_{jl} - q_l^2\,\omega_{ij}\right) + q_l^2\,\omega_{ik}\,\omega_{jk}   \\[-3mm] 
                    & \qquad\qquad\quad + \omega_{kl}\left( \omega_{ij}\,\omega_{kl} - \omega_{ik}\,\omega_{jl} - \omega_{il}\,\omega_{jk}\right)\Big] \\
       &+ P^2 \sum_{i<j} \omega_{ij}^2 \,\left( q_k^2\,q_l^2 - \omega_{kl}^2\right) = 0\,,
   \end{split}
\end{equation}
where the indices $\{i,j,k,l\}$ in the sums are all different.
For example, if all $\omega_{ij}$ vanish, one arrives at
\begin{equation}
   \begin{split}
      \sum_i \eta_i^2 \,q_j^2\,q_k^2\,q_l^2 &= \eta_1^2\,q_2^2\,q_3^2\,q_4^2 + \eta_2^2\,q_1^2\,q_3^2\,q_4^2 \\[-3mm]
                                            &+ \eta_3^2\,q_1^2\,q_2^2\,q_4^2 + \eta_4^2\,q_1^2\,q_2^2\,q_3^2 \\
                                            &= q_1^2\,q_2^2\,q_3^2\,q_4^2\,P^2\,,
   \end{split}
\end{equation}
which implies that not all $\eta_i$ can vanish simultaneously (unless one of the $q_i^2$ is zero).
   
We  note that the constraint~\eqref{dim-constraint} can also be expressed through the permutation-group multiplets given in Eq.~\eqref{5b-singlet} and below,
but this leads to a complicated  relation involving all quintic expressions of the form $(\mQ_1\cdot \mQ_1)^2 \,P^2$, $(\mQ_1\cdot \mQ_1)(\mV\cdot\mV)\,P^2$,
$(\mQ_1\vee\mQ_1)\cdot(\mV\vee\mV)\,P^2$, $\left[ (\mQ_1\cup\mV)\vee\mV\right]\cdot (\mQ_2\vee\mQ_2)$, etc.

\newpage

\bibliography{lit}

\end{document}